\documentclass[onecolumn,10pt,journal]{IEEEtran}
\usepackage{times}
\usepackage{amsmath}  
\usepackage{amssymb}  
\usepackage{mathrsfs} 
\usepackage{theorem}  
\usepackage{cite}     
\usepackage{comment}  
\usepackage{pifont}
\usepackage{amsfonts}
\usepackage{verbatim}
\usepackage[dvipsnames,usenames]{color}
\usepackage{enumerate}
\usepackage{graphicx}
\usepackage{subfigure}
\usepackage{latexsym}
\usepackage{caption}
\usepackage{algorithm}
\usepackage[noend]{algorithmic}
\usepackage{psfrag}
\usepackage[normalem]{ulem}
\usepackage{stmaryrd}
\usepackage{amscd}
\usepackage{color}
\usepackage{times,color}
\usepackage[usenames,dvipsnames,svgnames,table]{xcolor}

\usepackage[top=0.56in, bottom=0.56in, left=0.58in, right=0.58in]{geometry}

\newcommand{\be}[1]{\begin{equation}\label{#1}}
\newcommand{\ee}{\end{equation}}

\newcommand{\bc}{\begin{center}}
\newcommand{\ec}{\end{center}}

\newcommand{\qed}{\hfill$\Box$\\[1ex]}

\newcommand{\cC}{{\cal{C}}}
\newcommand{\cD}{{\cal D}}
\newcommand{\cE}{{\cal E}}

\newcommand{\cR}{{\cal R}}

\newcommand{\bfa}{{\boldsymbol a}}
\newcommand{\bfb}{{\boldsymbol b}}
\newcommand{\bfc}{{\boldsymbol c}}

\newcommand{\bfp}{{\boldsymbol p}}
\newcommand{\bfq}{{\boldsymbol q}}

\newcommand{\bfv}{{\boldsymbol v}}
\newcommand{\bfw}{{\boldsymbol w}}

\newcommand{\bfR}{{\mathbf R}}

\renewcommand{\le}{\leqslant}
\renewcommand{\leq}{\leqslant}
\renewcommand{\ge}{\geqslant}
\renewcommand{\geq}{\geqslant}

\newcommand{\B}{\mathcal B}

\newcommand{\D}{\mathcal D}
\newcommand{\E}{\mathcal E}

\newcommand{\Cref}[1]{Co\-rol\-la\-ry\,\ref{#1}}

\theoremstyle{plain} \theorembodyfont{\normalfont\slshape}

\newtheorem{thm}{Theorem$\!$}
\newenvironment{theorem}{\begin{thm}\hspace*{-1ex}{\bf.}}{\end{thm}}

\newtheorem{prop}[thm]{Proposition$\!$}
\newenvironment{proposition}{\begin{prop}\hspace*{-1ex}{\bf.}}{\end{prop}}

\newtheorem{lem}[thm]{Lemma$\!$}
\newenvironment{lemma}{\begin{lem}\hspace*{-1ex}{\bf.}}{\end{lem}}

\newtheorem{cor}[thm]{Corollary$\!$}
\newenvironment{corollary}{\begin{cor}\hspace*{-1ex}{\bf.}}{\end{cor}}

\newtheorem{const}[thm]{Construction$\!$}
\newenvironment{construction}{\begin{const}\hspace*{-1ex}{\bf.}}{\end{const}}

\newtheorem{proc}[thm]{Procedure$\!$}

\newtheorem{cl}[thm]{Claim$\!$}

\newtheorem{conject}[thm]{Conjecture$\!$}

\newtheorem{defi}[thm]{Definition$\!$}
\newenvironment{definition}{\begin{defi}\hspace*{-1ex}{\bf.}}{\end{defi}}

\theorembodyfont{\normalfont}

\newtheorem{exam}{Example$\!$}
\newenvironment{example}{\begin{exam}\hspace*{-1ex}{\bf .}}{\qed\end{exam}}

\newtheorem{remrk}{Remark$\!$}
\newenvironment{remark}{\begin{remrk}\hspace*{-1ex}{\bf .}}{\end{remrk}}

\definecolor{Codecolor}{named}{White}

\newcommand{\Copen}{\mbox{\{\kern-5.50pt\{}}
\newcommand{\Cclose}{\mbox{\}\kern-5.50pt\}}}
\newcommand{\Cslash}{\mbox{$\backslash\kern-6.02pt\backslash$}}

\newtheorem{innercustomgeneric}{\customgenericname}
\providecommand{\customgenericname}{}
\newcommand{\newcustomtheorem}[2]{%
	\newenvironment{#1}[1]
	{%
		\renewcommand\customgenericname{#2}%
		\renewcommand\theinnercustomgeneric{##1}%
		\innercustomgeneric
	}
	{\endinnercustomgeneric}
}
\newcustomtheorem{customlemma}{Lemma}
\newcustomtheorem{customthm}{Theorem}

\begin{document}
\title{Endurance-Limited Memories: Capacity and Codes}

\author{\IEEEauthorblockN{Yeow Meng Chee, 
			\IEEEmembership{Senior Member IEEE},
			Michal Horovitz, 
			\IEEEmembership{Member IEEE},
			Alexander Vardy, 
			\IEEEmembership{Fellow IEEE},\\
			Van Khu Vu,
			and Eitan Yaakobi, 
			\IEEEmembership{Senior Member IEEE}	
			}
{
\thanks{Parts of the results in this paper were presented in the International Symposium on Information Theory and Its Applications, Oct 2018~\cite{CHVVY18}, and in IEEE Information Theory Workshop, August 2019~\cite{CHVVY19}.}
\thanks{Yeow Meng Chee is with the Department of Industrial Systems Engineering
and Management, National University of Singapore, Singapore	(email: ymchee@nus.edu.sg). }
\thanks{Michal Horovitz is with the Department of
		Computer Science,
		Tel-Hai College,
		and The Galilee Research Institute - Migal, Kiryat Shmona 11016, Upper Galilee Israel 
		(e-mail: horovitzmic@telhai.ac.il).}
\thanks{Alexander Vardy is with the Department of Electrical and Computer Engineering
and the Department of Computer Science and Engineering, University of
California San Diego, La Jolla, CA 92093 USA (e-mail: avardy@ucsd.edu).}
\thanks{ Van Khu Vu is with the Department of Industrial Systems Engineering
and Management, National University of Singapore, Singapore (email: isevvk@nus.edu.sg). }
\thanks{Eitan Yaakobi is with the Computer Science  Department, Technion --- Israel Institute of Technology, Haifa 3200003, Israel (e-mail: yaakobi@cs.technion.ac.il).}
}
}

\maketitle

\thispagestyle{empty}
\pagestyle{empty}

\begin{abstract}
\emph{Resistive memories}, such as \emph{phase change memories} and \emph{resistive random access memories} have attracted significant attention in recent years due to their better scalability, speed, rewritability, and yet non-volatility. However, their \emph{limited endurance} is still a major drawback that has to be improved before they can be widely adapted in large-scale systems.

In this work, in order to reduce the wear out of the cells, we propose a new coding scheme, called \emph{endurance-limited memories} (\emph{ELM}) codes, that increases the endurance of these memories by limiting the number of cell programming operations. Namely, an \emph{$\ell$-change $t$-write ELM code} is a coding scheme that allows to write $t$ messages into some $n$ binary cells while guaranteeing that each cell is programmed at most $\ell$ times. In case $\ell=1$, these codes coincide with the well-studied \emph{write-once memory} (\emph{WOM}) codes. We study some models of these codes which depend upon whether the encoder knows on each write the number of times each cell was programmed, knows only the memory state, or even does not know anything. For the decoder, we consider these similar three cases. 
We fully characterize the capacity regions and the maximum sum-rates of three models where the encoder knows on each write the number of times each cell was programmed. In particular, it is shown that in these models the maximum sum-rate is $\log \sum_{i=0}^{\ell} {t \choose i}$. We also study and expose the capacity regions of the models where the decoder is informed with the number of times each cell was programmed. Finally we present the most practical model where the encoder read the memory before encoding new data and the decoder has no information about the previous states of the memory.
\end{abstract}

\section{Introduction}
Emerging resistive memory technologies, such as \emph{resistive random access memories} (\emph{ReRAM}) and \emph{phase-change memories}
(\emph{PCM}), have the potential to be the future's universal memories. They combine several important attributes starting from the speed of SRAM, the density of DRAM, and the non-volatility of flash memories. However, they fall short in their \emph{write endurance}, which significantly increases their bit error rate (BER). Hence, solving the limited endurance of these memories is crucial before they can be widely adapted in large-scale systems \cite{Getal19, Setal18, Zhaoetal18}.

Resistive memories are non-volatile memories which are composed of cells. The information is stored in the cells by changing their resistance. They combine the following two properties of DRAM and flash memories. Similarly to flash memories and unlike DRAM they are non-volatile memories and thus they don't require refresh operations. Furthermore, like DRAM and unlike flash memories they are rewritable without an erase operation. The main challenge that has remained to be solved in order to make these memories a legitimate candidate as a universal memory is their limited write endurance, which is the goal of this paper. 

Endurance is defined as the number of set/reset cycles to switch the state of cells in ReRAM while it is still reliable. Owing to its importance, there are many researches which test and characterize the endurance of ReRAM in order to show a strong dependence on material of cells, cell size, \cite{Ranaetal17, ZAK20}, and program operation \cite{Wangetal15}. To improve the endurance of ReRAM, recent research have focused on the structure and material of devices \cite{Ranaetal17, Yuanetal17} and programming schemes \cite{Chenetal11, Wangetal15}. In this work, we present a scheme to use rewriting code to improve the endurance lifetime of ReRAM.

Previous works have offered different solutions to combat the write endurance of resistive memories. In~\cite{Ketal16}, the authors proposed to use locally repairable codes (LRC) in order to construct codes with small rewriting locality in order to mitigate both the problems of endurance and power consumption. In~\cite{Zetal16}, the authors proposed mellow writes, a technique which is targeted to reduce the wear-out of the writes rather than reducing the number of writes. Lastly, several other works proposed coding schemes which correct stuck-at cells; see e.g.~\cite{MMC15,SLB10,XNZYX15}.

In order to combat the limited write endurance in resistive memories, this paper proposes to study the following new family of codes, called \emph{endurance-limited memory} (\emph{ELM}) codes. Assume there are $n$ binary cells and $t$ messages that are required to be stored in these cells sequentially. Assume also that each cell can be programmed at most $\ell\geq 1$ times. Then, we seek to find the set of achievable rates, i.e., the capacity region, and design code constructions for this model. Note that for $\ell=1$, we get the classical problem of write-once memory (WOM) codes~\cite{CKVY20, H85,HY17,RS82,Sh14,WWZK84}. Besides that, if $t \leq \ell,$ the coding scheme is trivial. Hence, in this work, we only focus on the cases where $t > \ell >1$. 
Note that a trivial lower bound on the maximum sum-rate is $\min\{t,\ell\}$, which achieved by writing with rate
$1$ in the first $\min\{ t,\ell\}$ writes and with rate $0$ in the remaining writes.

Let us consider first the case where $\ell=2$ and $t=3$. A naive solution is to use a two-write WOM code for the first two writes and then write $n$ more bits on the third write. The maximum sum-rate using this solution will be $\log(3) +1 = \log(6)$, while, as will be shown in the paper, the maximum sum-rate in this case is $\log(7)$. 
The intuition behind this is as follows. Let $p_1$ be the probability to program a cell on the first write, so we assume that $p_1n$ cells are programmed. Then, on the second and third writes we have a two-write WOM code problem for the $p_1n$ programmed cells, and for the $(1-p_1)n$ non-programmed cells, we can write twice on them so no coding is needed. The maximum sum-rate is achieved for $p_1=3/7$.
However, it is still a challenging task to design specific code constructions that can approach sum-rate of $\log(7)$.

There are several models of ELM codes which can be studied. These models are distinguishable by the information that is available to the encoder and the decoder. In particular, for the encoder, we consider three cases which depend upon whether the encoder knows the number of times each cell was programmed, \emph{encoder informed all} (\emph{EIA}), only the current state of the cell, \emph{encoder informed partially} (\emph{EIP}), or no information about the cells state, \emph{encoder uninformed} (\emph{EU}). The decoder will also have three cases, corresponding to the same information that is available to the encoder. 
Thus, by considering all combinations of the above three cases for the encoder and the decoder, it is possible to define and study nine models, $EX:DY$, where $X,Y\in\{IA,IP,U\}$.
The rest of this paper is organized as follows. 
In Section~\ref{sec:def}, we formally define the models studied in this paper and discuss some basic observations. In Section~\ref{sec:EIA}, we study the capacity regions and the maximum sum-rates of the EIA models, and also present capacity achieving codes. We prove that the capacity region of all EIA models, are the same, for both $\epsilon$-error and zero-error cases. 
In the next two sections, we discuss the EIP:DIA model.
In Section~\ref{sec:EIP:DIA}, we study the capacity region of this model for the $\epsilon$-error case, 
and in Section \ref{sec:comparison:EIAvsEIP:DIA}, we compare between this model and the EIA models.
Then, we discuss the EU:DIA and the EIP:DU models in
Sections~\ref{sec:EU:DIA} and~\ref{sec:EIP:DU}, respectively. Finally, we conclude our results and discuss a future work in Section~\ref{SEC.conclusion}.

\section{Definitions and Preliminaries}\label{sec:def}

In this section, we formally define the nine models of ELM codes,
and we state some simple propositions.
Assume that each cell can be programmed at most $\ell$ times, so if the encoder attempts to program a cell more than $\ell$ times then it will not change its value. 
For the EIA models, we assume that the encoder will not try to program a cell that has already been programmed $\ell$ times before the current write. We see this as an extension of the WOM model for $\ell=1$. 
These models will be defined both for the zero-error and the $\epsilon$-error cases.

For a positive integer $a$, the set $\{0,\ldots,a-1\}$ is defined by~$[a]$. Throughout this paper, we assume that the number of cells is~$n$. We use the vector notation $\bfc\in[2]^n$ to represent the \emph{cell-state vector} of the $n$ memory cells, and the vector $\bfv \in [\ell+1]^n$, which will be called the \emph{cell-program-count vector}, to represent the number of times each cell was programmed. Note that the state of a cell is the parity of the number of times it was programmed. Thus, if the encoder (or the decoder) knows the cell-program-count vector $\bfv$, in particular it knows the cell-state vector $\bfc$ as well.
For a vector $\bfv \in [\ell+1]^n$, we denote by ${\langle \bfv \rangle}_2$ the length-$n$ binary vector which satisfies  ${\langle \bfv \rangle}_{2,k}=\bfv_k(\bmod\ 2)$ for all $k\in [n]$, and we say that ${\langle \bfv \rangle}_2$ equals to $\bfv$ modulo $2$.
The complement of a binary vector $\bfc$ is denoted by $\overline{\bfc}$. The all ones, zeros vector will be denoted by $\textbf{1}$, $\bf0$, respectively.
For two length-$n$ vectors $\bfa$ and $\bfb$,
$\bfa+\bfb$ is the vector obtained by pointwise addition.
If $\bfa$ and $\bfb$ are binary vectors
$\bfa\oplus\bfb$ is the vector obtained by pointwise addition modulo $2$.

For a cell-program-count vector $\bfv \in[\ell+1]^n$ and a new cell-state vector $\bfc \in [2]^n$ to be programmed to the cells, we define by $N(\bfv,\bfc)\in[\ell+1]^n$ and $f(\bfv,\bfc)\in [2]^n$ the result of programming the new cell-state vector $\bfc$. That is, $N(\bfv,\bfc)$ is the new cell-program-count vector after programming $\bfc$, $f(\bfv,\bfc)$ is the new cell-state vector, and they are formally defined as follows. 
$N(\bfv,\bfc)_k=\bfv_k$ if $\bfc_k=\bfv_k (\bmod \ 2)$, and otherwise 
$N(\bfv,\bfc)_k=\min \{\ell,\bfv_k+1\}$, where the index $k$ for a vector means its $k$-th element. Similarly, $f(\bfv,\bfc)_k=\bfc_k$ if $\bfv_k<\ell$, and otherwise $f(\bfv,\bfc)_k=\bfv_k(\bmod\ 2)$.
Note that ${\langle N(\bfv,\bfc)\rangle}_2=f(\bfv,\bfc)$, i.e., $f(\bfv,\bfc)$ equals to 
$N(\bfv,\bfc)$ modulo~$2$.
We are ready now to define all models studied in this paper.
\begin{definition}
\label{def:ELM_code}
An $[n,t,\ell;M_1,\dots, M_t]^{EX:DY,\bfp_e}$ $\ell$-change $t$-write \textbf{Endurance-Limited Memory} (\textbf{ELM}) code with error-probability vector $\bfp_{e}=(p_{e_1},\ldots,p_{e_t})$, where $X,Y \in \{IA,IP,U\}$, is a coding scheme comprising of $n$ binary cells and is defined by $t$ encoding and decoding maps $(\E_j, \D_j)$ for $1 \leq j \leq t$. For the map $\E_j$, $Im(\E_j)$ is its image, where by definition $Im(\E_0)=\{(0,\ldots,0)\}$.

Furthermore, for $j\in [t+1]$, let $N_j$ and $Im^*(\E_j)$ be the sets of all state-program-count vectors, cell-state vectors which can be obtained after the first $j$ writes, respectively.
Formally, for ${1\leq j}$, $N_j =\{N(\bfv,\bfc) :\bfc\in Im(\E_j), \bfv\in N_{j-1}\}$, where $N_{0}=\{(0,\ldots,0)\}$,
and 
$Im^*(\E_j)=\{{\langle \bfv \rangle}_2 : \bfv \in N_{j}\}.$
Note that for the EIA models $Im^*(\E_j)=Im(\E_j)$.
The domain and the range of the encoding maps are defined as follows:
\begin{enumerate}[(1)]
\item for the EIA models,
$\E_j:[M_j] \times N_{j-1} \to [2]^n,$
such that for all $(m,\bfv)\in [M_j] \times N_{j-1}$ it holds that
$\bfv+\left({\langle \bfv \rangle}_2 \oplus \E_j(m,\bfv)\right)\in [\ell+1]^n$.
\item for the EIP models,
$\E_j:[M_j] \times Im^{*}(\E_{j-1}) \to [2]^n$.
\item for the EU models,
$\E_j:[M_j] \to [2]^n.$
\end{enumerate}

For a message $m$ we denote by $I_m(x)$ the indicator function, where $I_m(x)=0$ if $m=x$, and otherwise $I_m(x)=1$.
Additionally, for a message $m\in [M_j]$, $Pr(m)$ is the probability of programming a message $m$ on the $j$-th write, and for $\bfv \in N_{j-1}$, $Pr(\bfv)$ is the probability to have cell-program-count vector $\bfv$ before the $j$-th write.
The nine models are defined as follows.
For all $1\leq j\leq t$,

\begin{enumerate}[(1)]

\item \label{EIA:DIA} if $(X,Y)= (IA,IA)$ then 
$$
\D_j:  \{(\E_j(m,\bfv),\bfv) : m\in[M_j],\bfv\in N_{j-1}\} \hspace*{-0.1cm} \to \hspace*{-0.1cm} [M_j],$$
$$\sum_{(m,\bfv)\in [M_j] \times N_{j-1}}\hspace*{-6ex}Pr(m)Pr(\bfv)\cdot 	I_m\left(\D_j(\E_j(m,\bfv),\bfv)\right) \leq 	p_{e_j},$$

\item \label{EIA:DIP} if $(X,Y)= (IA,IP)$ then
$$\D_j:  \{(\E_j(m,\bfv),{\langle \bfv \rangle}_2) : m\in[M_j], \bfv\in N_{j-1}\} \hspace*{-0.1cm} \to \hspace*{-0.1cm} [M_j],$$
$$\sum_{(m,\bfv)\in [M_j] \times N_{j-1}}\hspace*{-6ex}Pr(m)Pr(\bfv)\cdot 	I_m\left(\D_j(\E_j(m,\bfv),{\langle \bfv \rangle}_2)\right) \leq 	p_{e_j},$$

\item \label{EIA:DU} if $(X,Y)= (IA,U)$ then 
$$\D_j:  Im(\E_j) \to \hspace*{-0.1cm} [M_j],$$
$$\sum_{(m,\bfv)\in [M_j] \times N_{j-1}}\hspace*{-6ex}Pr(m)Pr(\bfv)\cdot 	I_m\left(\D_j(\E_j(m,\bfv))\right) \leq 	p_{e_j},$$
\item \label{EIP:DIA} if $(X,Y)= (IP,IA)$ then
$$ \hspace{-0.7ex}\D_j:  \{(f(\bfv,\E_j(m,{\langle \bfv \rangle}_2)),\bfv) : m\in[M_j],\bfv\in N_{j-1}\} \hspace*{-0.1cm} \to \hspace*{-0.1cm} [M_j],$$ 
$$\hspace*{-3.6ex}\sum_{(m,\bfv)\in [M_j] \times N_{j-1}}\hspace*{-6ex}Pr(m)Pr(\bfv)\cdot 	I_m\left(\D_j(f(\bfv,\E_j(m, {\langle \bfv \rangle}_2)),\bfv)\right) \leq 	p_{e_j},$$

\item \label{EIP:DIP} if $(X,Y)= (IP,IP)$ then 
$$ \hspace{-0.7ex}\D_j:  \{(f(\bfv,\E_j(m,{\langle \bfv \rangle}_2)),{\langle \bfv \rangle}_2) : m\in[M_j],\bfv\in N_{j-1}\} \hspace*{-0.1cm} \to \hspace*{-0.1cm} [M_j],$$ 
$$\hspace*{-3.6ex}\sum_{(m,\bfv)\in [M_j] \times N_{j-1}}\hspace*{-6ex}Pr(m)Pr(\bfv)\cdot 	I_m\left(\D_j(f(\bfv,\E_j(m, {\langle \bfv \rangle}_2)),{\langle \bfv \rangle}_2)\right) \leq 	p_{e_j},$$

\item \label{EIP:DU} if $(X,Y)= (IP,U)$ then 
$$\D_j:  Im^{*}(\E_j) \to \hspace*{-0.1cm} [M_j],$$
$$\sum_{(m,\bfv)\in [M_j] \times N_{j-1}}\hspace*{-6ex}Pr(m)Pr(\bfv)\cdot 	I_m\left(\D_j(f(\bfv,\E_j(m, {\langle \bfv \rangle}_2)))\right) \leq 	p_{e_j},$$

\item \label{EU:DIA} if $(X,Y)= (U,IA)$ then 
$$\D_j:  \{(f(\bfv,\E_j(m)),\bfv) : m\in[M_j],\bfv\in N_{j-1}\} \hspace*{-0.1cm} \to \hspace*{-0.1cm} [M_j],$$
$$\sum_{(m,\bfv)\in [M_j] \times N_{j-1}}\hspace*{-6ex}Pr(m)Pr(\bfv)\cdot 	I_m\left(\D_j(f(\bfv,\E_j(m))\bfv)\right) \leq 	p_{e_j},$$

\item \label{EU:DIP} if $(X,Y)= (U,IP)$ then
$$\D_j:  \{(f(\bfv,\E_j(m)),{\langle \bfv \rangle}_2) : m\in[M_j], \bfv\in N_{j-1}\} \hspace*{-0.1cm} \to \hspace*{-0.1cm} [M_j],$$
$$\sum_{(m,\bfv)\in [M_j] \times N_{j-1}}\hspace*{-6ex}Pr(m)Pr(\bfv)\cdot 	I_m\left(\D_j(f(\bfv,\E_j(m)){\langle \bfv \rangle}_2)\right) \leq 	p_{e_j},$$

\item \label{EU:DU} if $(X,Y)= (U,U)$ then 
$$\D_j:  Im^{*}(\E_j) \to \hspace*{-0.1cm} [M_j],$$
$$\sum_{(m,\bfv)\in [M_j] \times N_{j-1}}\hspace*{-6ex}Pr(m)Pr(\bfv)\cdot 	I_m\left(\D_j(f(\bfv,\E_j(m)))\right) \leq 	p_{e_j}.$$

\end{enumerate}
If $p_{e_j}=0$ for all $1\leq j\leq t$, then the code is called a zero-error ELM code and is denoted by $[n,t,\ell;M_1,\ldots,M_t]^{EX:DY,z}$.
\end{definition}

The \emph{rate} on the $j$-th write of an $[n,t,\ell; M_1,\dots, M_t]^{EX:DY,\bfp_{e}}$ ELM code, $X,Y \in \{IA,IP,U\}$, is defined as $R_j=\frac{\log M_j}{n}$, and the \emph{sum-rate} is the sum of the individual rates on all writes, $R_{sum}=\sum_{j=1}^{t} R_j$. A rate tuple $\bfR=(R_1,\ldots, R_t)$ is called \emph{$\epsilon$-error achievable} in model $EX:DY$, if for all $\epsilon >0$ there exists an $[n,t,\ell;M_1,\ldots,M_t]^{{EX:DY},\bfp_{e}}$ ELM code with error-probability vector $\bfp_e=(p_{e_1},\ldots,p_{e_t})\leq(\epsilon,\ldots,\epsilon)$, such that $\frac{\log M_j}{n}\ge R_j-\epsilon$. The rate tuple $\bfR$ will be called \emph{zero-error achievable} if for all $1\leq j\leq t$, $p_{e_j}=0$. The \emph{$\epsilon$-error capacity region} of the EX:DY model is the set of all $\epsilon$-error achievable rate tuples, that is, $$\cC^{EX:DY,\epsilon}_{t,\ell}=\{(R_1,\dots,R_t)|\text{$(R_1,\ldots,R_t)$ is $\epsilon$-error achievable}\},$$ and the \emph{$\epsilon$-error maximum sum-rate} will be denoted by $\cR^{EX:DY,\epsilon}_{t,\ell}$. The \emph{zero-error capacity region}  $\cC_{t,\ell}^{EX:DY,z}$ and the \emph{zero-error maximum sum-rate} $\cR_{t,\ell}^{EX:DY,z}$ are defined similarly.  We say that $\bfR \le \bfR'$ for $\bfR=(R_1,\ldots, R_t)$  and $\bfR'=(R_1',\ldots, R_t')$ if $R_j \le R_j'$ for all $1\le j \le t$,
and $\bfR < \bfR'$
if $\bfR \le \bfR'$ and $\bfR \ne \bfR'$.

According to these definitions it is easy to verify the following relations.
For $g\in \{z,\epsilon\}$ and $X,Y\in \{IA,IP,U\}$ it holds that
\begin{gather*}
\cC^{EU:DY,g}_{t,\ell} \subseteq 
\cC^{EIP:DY,g}_{t,\ell} \subseteq \cC^{EIA:DY,g}_{t,\ell}, \\
\cC^{EX:DU,g}_{t,\ell} \subseteq 
\cC^{EX:DIP,g}_{t,\ell} \subseteq \cC^{EX:DIA,g}_{t,\ell}, \text{and} \\
\cC^{EX:DY,z}_{t,\ell} \subseteq \cC^{EX:DY,\epsilon}_{t,\ell}.
\end{gather*}
Similar connections hold for the maximum sum-rates.

Note that if $\ell\geq t$ then all problems are trivial since it is possible to program all cells on each write, so the capacity region in all models is $[0,1]^t$ and the maximum sum-rate is $t$. 
For $\ell=1,$ we get the classical and well-studied WOM codes~\cite{CKVY20,H85,HY17,RS82,Sh14,WWZK84}. In this case, we also notice that the IA and IP models are the same for both the encoder and the decoder. The capacity region and the maximum sum-rate in most of these cases are known; see e.g.~\cite{H85,HY17, RS82,WWZK84}. In the rest of this paper, and unless stated otherwise, we assume that $1\leq \ell < t$. 
\subsection{Related Work}
The EIA models of ELM codes studied in this paper are strongly related to non-binary WOM codes and their modified versions studied in \cite{FH99,HY17, KMM18, CKVVY19}.
In these EIA models, we can treat every cell as an $(\ell+1)$-ary cell, where it is only possible to increase its level by one on each write, while its maximum level is $\ell$. In non-binary WOM codes, each cell has $q$ levels and its level can not be decreased\cite{FH99, HY17}. In a write $\ell$-step-up memory\cite{CKVVY19}, a special version of the non-binary WOM code, each cell has $q$ levels and each time we write a cell, its level can be increased by at most some value $\ell$. Recently, Kobayashi et al. \cite{KMM18} also studied a modified version of non-binary WOM codes, called write constrained memories, where there is a cost on each state transition. Yet, these codes are not identical and we can not apply previous results to solve the models of ELM codes. In fact, our results on EIA models are useful to obtain an explicit formula of the capacity region of write $\ell$-step-up memory codes. We also note that some models of ELM codes, such as the EIP:DU model, are much different from previous models and are difficult to solve. 

Moreover, our proposed ELM coding scheme is also related to the cooling code which is used to control the peak temperature of an interconnect\cite{CEKV18, CKVY20}. In \cite{CEKV18}, cooling codes are proposed to avoid all hottest wires and in \cite{CKVY20}, cooling codes are shown to be equivalent to two-write binary WOM codes. Later in this work, we will use cooling codes as two-write binary WOM codes in order to construct our ELM codes. Now, we discuss the ability of our ELM codes to control the peak temperature of an interconnection. In fact, using our coding scheme in this work, we can control the maximal number of switches in each wire. We note that the temperature of each wire is closely related to the number of switches in each wire. And thus, we can control the peak temperature of each wire.

Recently, the EIA:DIA model of ELM  is shown to be useful for two-dimensional (2D) weight-constrained code scheme~\cite{NVK20}. In a 2D weight-constrained code, each codeword is an array of size $m \times n$ where the number of 1 symbols in each row and each column is limited by $pn$ and $qm$, respectively. The encoder and decoder know the values of all the $mn$ bits. We can view this 2D weight-constrained code as we write $m$ messages in ReRAM where each cell can be switched at most $qm$ times and both encoder and decoder know all previous messages.

\subsection{Our Contribution}
In this work, we propose a novel scheme of rewriting code to improve the endurance lifetime of ReRAM, called endurance limited memories (ELM) code. In $\ell$-change $t$-write ELM codes, each cell can be programmed at most $\ell$ times during the writing of $t$ messages. In the case that $\ell$ is much smaller than $t$, we can significantly improve the lifetime of the memories. Depending upon whether the encoder (E) and decoder (D) know the number of times each cell is programmed (IA), the current state of each cell (IP), or have no information on the state of each cell (U), we present and investigate all nine models $EX:DY$ where $X,Y\in \{IA,IP,U\}$. We note that the most practical model to increase the endurance of ReRAM is the EIP:DU model. However, for the theoretical interest, we study all nine models in this paper. Furthermore, although the EIA models are not suitable for improving the endurance of ReRAM, they have several applications, including write $\ell$-step-up memories and two-dimensional weight-constrained codes. We expect that we can see other applications of all nine models of ELM codes in near future. 

In Section \ref{sec:EIA}, all three EIA models are investigated in both cases, $\epsilon$-error and zero-error. We first provide the capacity region and the maximum sum-rate of these models. The techniques are used to achieve the results on the capacity regions in Theorems~ \ref{thm:capacityEIA} and~\ref{thm:equiv} are similar to those used in \cite{HY17} for WOM codes. To achieve the explicit formula of the maximum sum-rate in Theorem~ \ref{thm:EIA_maxSumRate}, we need a new simple technique. We then present several constructions of ELM codes for the EIA models.  To construct these codes, we need some new ideas even though we also use several special families of WOM codes as components of our ELM codes. In Section \ref{sec:EIP:DIA}, we use a known technique in information theory to obtain the capacity region of the EIP:DIA model. The capacity region of all EIA models is compared to the EIP:DIA model in Section \ref{sec:comparison:EIAvsEIP:DIA}. Using the same technique carefully, we can also find the capacity region of the EU:DIA model in Section \ref{sec:EU:DIA}. Finally, in Section \ref{sec:EIP:DU}, we study the most practical model for ReRAM, the EIP:DU model. Although several good bounds on the maximum sum-rate are presented, these bounds are not tight and finding an exact formula of the maximum sum-rate of the EIP:DU model is still an open problem. To achieve some good constructive lower bounds, we provide several constructions of EIP:DU ELM codes for the zero-error case. These results are novel and we need different original methods to achieve them.
\section{The EIA Models}\label{sec:EIA}
In this section, we explore the capacity region and the maximum sum-rate of the EIA models for both the $\epsilon$-error and the zero-error cases.
We also propose capacity achieving codes for these cases. 

For each $j\in[t+1]$, we let $\bfc_j$ denote the binary length-$n$ vector which represents the cell-state vector after the $j$-th write, where $\bfc_0=\bf0$. 
Recall that in the EIA models, on the $j$-th write the encoder knows the number of times each cell was programmed before the current write. That is, the encoder receives as an input a length-$n$ cell-program-count vector $N(\bfc_{j-1})\in [\ell+1]^n$ that represents the number of times each cell was programmed so far. 
Next, for all $t$ and $\ell,$ we define the region $\cC_{t,\ell}$ and in Theorem~\ref{thm:capacityEIA}, we prove that this is the capacity region of all the EIA models.

For $1\leq j\leq t$ and $i\in[\ell+1]$, $i\leq j$, let $p_{j,i}\in[0,0.5]$, be the probability to program a cell on the $j$-th write, given that this cell has been already programmed $i$ times. 
We define $p_{j,j}=p_{j,\ell}=0$ for $1\leq j \leq t$,
and let $Q_{j,i}$ be the probability that a cell has been  programmed exactly $i$ times on the first $j$ writes. Formally, $Q_{j,i}$ is defined recursively by using $p_{j,i}$ and $p_{j,i-1}$ as follows:
\begin{equation}
\label{eq:EIA:DIqProbFunc}
\begin{array}{rl}
\hspace*{-0.1cm}Q_{j,i}\hspace*{-0.2cm}&=\begin{cases}
Q_{j-1,i}(1-p_{j,i})+Q_{j-1,i-1}p_{j,i-1}, & \text{if } i>0,\\
Q_{j-1,i}(1-p_{j,i}), & \text{if } i=0,
\end{cases}\\
\end{array}
\end{equation}
where for $j=0$ we set
$Q_{0,0}=1$ otherwise $Q_{0,i}=0$.
The rates region $\cC_{t,\ell}$ is defined as follows:
\begin{equation}
\label{eq:capacityModelEIA}
\begin{array}{ll}
\hspace{-2ex} \cC_{t,\ell}\hspace{-0.5ex}= 
\hspace{-0.5ex}
\Big\{ \hspace{-0.5ex}(R_1,\ldots,R_t)|
\forall 1\hspace{-0.5ex}\leq\hspace{-0.5ex} j\hspace{-0.5ex}\leq\hspace{-0.5ex} t:  R_j \leq \sum_{i=0}^{\min\{\ell,j\}-1}Q_{j-1,i} h(p_{j,i}),\\
 \hspace{8ex} \forall  i\in[\ell]: p_{j,i} \in [0,0.5], 
\text{ and} \ Q_{j,i} \text{ is defined in (\ref{eq:EIA:DIqProbFunc})} \Big\},
\end{array}
\end{equation}
where in this paper $h(x), H(X)$
is the binary entropy function where
$0\leq x \leq 1$, $X$ is a random variable, respectively.

Note that for $\ell=1$, it is possible to verify that we get the capacity region of WOM~\cite{H85,RS82,WWZK84}. 
It is also readily verified that 
the maximum sum-rate is achieved with $p_{j,i}=0.5$ for all $t-j+1\ge \ell-i$, since $t-j+1$ is the number of the remaining writes, and $\ell-i$ is the number of times the cell can be programmed. Thus, if $t-j+1\ge \ell-i$ then we can program the cell with probability $0.5$ to obtain the maximum rate.
The next theorem proves that for $2\leq \ell \leq t-1$, $\cC_{t,\ell}$ is the 
capacity region of the $\ell$-change $t$-write ELM in all EIA models,
and thus we denote this capacity by $\cC_{t,\ell}^{EIA}$, and the maximum sum-rate by $\cR_{t,\ell}^{EIA}$.
\begin{theorem}
	\label{thm:capacityEIA}
	The rates region $\cC_{t,\ell}$ is the capacity region of the $\ell$-change $t$-write ELM in all EIA models for both $\epsilon$-error and zero-error cases. 
    That is,
    for all $g\in \{z,\epsilon\}$ and $Y\in \{IA,IP,U\}$,
    $\cC_{t,\ell}^{EIA:DY,g} = \cC_{t,\ell}.$
\end{theorem}
\begin{IEEEproof}
Recall that by the definitions of the models
\begin{gather*}
\cC^{EIA:DU,z}_{t,\ell} \subseteq 
\cC^{EIA:DIP,z}_{t,\ell} \subseteq 
\cC^{EIA:DIA,z}_{t,\ell} \subseteq 
\cC^{EIA:DIA,\epsilon}_{t,\ell} , \text{and}\\
\cC^{EIA:DU,z}_{t,\ell} \subseteq 
\cC^{EIA:DU,\epsilon}_{t,\ell} \subseteq 
\cC^{EIA:DIP,\epsilon}_{t,\ell} \subseteq 
\cC^{EIA:DIA,\epsilon}_{t,\ell}.
\end{gather*}
The rest of the proof consists of two parts. The first part, called the direct part, proves that $\cC_{t,\ell}\subseteq \cC_{t,\ell}^{EIA:DU,z}$, and in the second, called the converse part, we prove that $\cC_{t,\ell}^{EIA:DIA,\epsilon} \subseteq \cC_{t,\ell}$. 
The direct part is proved in Subsection~\ref{subsec:EIA:DU}
for the zero-error case of the EIA:DU model, while the 
converse part is proved in Subsection~\ref{subsec:EIA:DIA}
for the $\epsilon$-error case of the EIA:DIA model.
\end{IEEEproof}

Next, we seek to present the capacity region of the EIA models in a recursive form. While we see this representation of the capacity region more intuitive, it will also help us in finding the maximum sum-rate of this model.
For all $t\ge 1$ and $\ell \ge 1$, let ${\widehat{\cC}}_{t,\ell}$ be the following region which is defined recursively as follows. For $t>\ell\ge 1$
\begin{align}\label{eq:capacityModelEIA:DIrecurssive}
\nonumber\hspace*{-1.5ex}\widehat{\cC}_{t,\ell}\hspace*{-0.5ex}=\hspace*{-0.5ex}\Big\{&(R_1,\ldots,R_t)|  R_1\le h(p), p\in [0,0.5], \\ 
\nonumber
& \text{for } 2 \leq j \leq t, R_j \le p R_j' + (1-p) R_j'', \\ 
& (R_2',\ldots,R_t') \in \widehat{\cC}_{t-1,\ell-1} \text{ and } (R_2'',\ldots,R_t'')\hspace*{-0.5ex} \in \hspace*{-0.5ex}\widehat{\cC}_{t-1,\ell}\Big\},
\end{align}
where for all $\ell\geq t \geq 1$ we set
$\widehat{\cC}_{t,\ell}= [0,1]^t$ and $\widehat{\cC}_{t,0}=\{\bf0\}$.
\begin{theorem}\label{thm:equiv}
For all $t$ and $\ell$, 
$\widehat{\cC}_{t,\ell}={\cC}_{t,\ell}$.
\end{theorem}
\begin{IEEEproof}
For the first direction, we prove by induction on $t$ that for all $\ell\geq 1$, 
if $\bfR=(R_1,\ldots,R_t)=\widehat{\cC}_{t,\ell}$ then 
$\bfR \in \cC^{EIA:DIA,\epsilon}_{t,\ell}$. 
Since $\cC^{EIA:DIA,\epsilon}_{t,\ell}={\cC}_{t,\ell}$,
we conclude that $\widehat{\cC}_{t,\ell} \subseteq {\cC}_{t,\ell}$.

The base of the induction is $t\leq \ell$ for all $\ell\ge 1$. These cases are readily verified.
For the induction step, let $\bfR=(R_1,R_2,\ldots,R_t)\in \widehat{\cC}_{t,\ell}$, $1\leq\ell<t$, 
such that 
$R_1= h(p)$ for $p \in [0,0.5]$ and for $2 \leq j \leq t$
$R_j = p  R_j' + (1-p) R_j''$ 
where
$(R_2',R_3',\ldots,R_t') \in \widehat{\cC}_{t-1,\ell-1}$ and
$(R_2'',R_3'',\ldots,R_t'') \in \widehat{\cC}_{t-1,\ell}$.
By the induction hypothesis, $(R_2',R_3',\ldots,R_t') \in \cC^{EIA:DI}_{t-1,\ell-1}$ and
$(R_2'',R_3'',\ldots,R_t'') \in \cC^{EIA:DI}_{t-1,\ell}$.
Thus, we have two codes: $C_1$ - an $(\ell-1)$-change $(t-1)$-write ELM code which achieves the rate tuple $(R_2',R_3',\ldots,R_t')$
and $C_2$ - an $\ell$-change $(t-1)$-write ELM code which achieves the rate tuple $(R_2'',R_3'',\ldots,R_t'')$.
Then, we can design an $\ell$-change $t$-write ELM code, such that on the first write
the encoder programs a cell with probability $p$ for $p \in [0,0.5]$,
and then on the next writes it applies $C_1$ for the cells that were programmed on the first write, and $C_2$ for the other cells.
Thus, the rate tuple $\bfR$ is achieved.

The second direction, ${\cC}_{t,\ell} \subseteq \widehat{\cC}_{t,\ell}$, is  proved by induction on $t$,
that is, for each $t \ge 1$ we prove that
${\cC}_{t,\ell}\subseteq \widehat{\cC}_{t,\ell}$ for all $1 \leq \ell \leq t$. 
The base of the induction, $t=1$ and $\ell =1$, is trivial.
The induction assumption is that for each $1 \leq \ell' \leq t-1$, 
${\cC}_{t-1,\ell'}\subseteq \widehat{\cC}_{t-1,\ell'}$.
For the induction step, let $\bfR=(R_1,R_2,\ldots,R_t)\in {\cC}_{t,\ell}$
which is achieved by the probabilities $p_{j,i}$.
Denote by 
$\bfR'=(R_2',R_3',\ldots,R_t')\in {\cC}_{t-1,\ell-1} $
the rate tuple which is attained by the probabilities $p'_{j,i}=p_{j+1,i+1}$,
and by
$\bfR''=(R_2'',R_3'',\ldots,R_t'')\in {\cC}_{t-1,\ell}$
the rate tuple which is attained by the probabilities $p''_{j,i}=p_{j+1,i}$.
Recall that we define 
$\widehat{\cC}_{t-1,t}=\widehat{\cC}_{t-1,t-1}$,
and $\widehat{\cC}_{t-1,0}=\{\bf0\}$.
It can be easily verified that
for all $j$, $2\leq j \leq t$,
$R_j=p_{1,0}R_j'+(1-p_{1,0})R_j''$.
By the induction hypothesis,
$\bfR' \in \widehat{\cC}_{t-1,\ell-1}$ and
$\bfR'' \in \widehat{\cC}_{t-1,\ell}$.
Thus, by defining $p=p_{1,0 }$ we get a recursive form for $\bfR$,
and we can conclude that
${\cC}_{t,\ell} \subseteq \widehat{\cC}_{t,\ell}$.
\end{IEEEproof}

Next, using the result from Theorem~\ref{thm:equiv}, it is possible to find the maximum sum-rate of the EIA models, $\cR^{EIA}_{t,\ell}$. 
\begin{theorem}\label{thm:EIA_maxSumRate}
For all $t$ and $\ell$,
$$\cR^{EIA}_{t,\ell}= \log \sum_{i=0}^{\ell} {t \choose i},$$
and this value is achieved for $$p_{1,0}=p=\frac{\sum_{i=0}^{\ell-1} {t-1 \choose i}}{\sum_{i=0}^{\ell} {t \choose i}},$$
where $p_{1,0}$, $p$ are defined in Equations~(\ref{eq:capacityModelEIA}),~(\ref{eq:capacityModelEIA:DIrecurssive}) in $\cC_{t,\ell}$, $\widehat{\cC}_{t,\ell}$, respectively.
For example, if $\ell=2$ the maximum sum-rate is achieved for $p_{1,0}=p=\frac{2t}{t^2+t+2}$.
\end{theorem}
\begin{IEEEproof}
First, we prove that $\cR^{EIA}_{t,\ell} \le \log \sum_{i=0}^{\ell} {\binom{t}{i}}$ by counting all the possible sequences of $t$ messages.
We describe each possible sequence as a table of $t$ rows and $n$ columns, where $n$ is the number of cells. Note that different sequences will be mapped to different matrices.
Recall, that every cell can be programmed at most $\ell$ times.
Thus, the number of different possible matrices is $\left(\sum_{i=0}^{\ell} {\binom{t}{i}}\right)^n,$ and the upper bound is proved.

Next we assure that this upper bound is indeed tight.
We prove this result by using the recursive formula for the capacity $\widehat{\cC}_{t,\ell}$ described in Equation~(\ref{eq:capacityModelEIA:DIrecurssive}).
For $\ell=1$, W$\ell$M is the binary WOM, and this upper bound, $\log(t+1)$ is known to be tight, and achieved for $p=1/(t+1)$~\cite{H85,FH99,WWZK84}.
That is, the maximum sum-rate of one-change $t$-write W$\ell$M is equal to $ \log \sum_{i=0}^{\ell} {t \choose i}=\log (t+1)$.
Let us denote $X_{t,\ell}=\sum_{i=0}^{\ell} {t \choose i}$ 
and $p=\frac{X_{t-1,\ell-1}}{X_{t,\ell}}$.
Note that by the properties of the binomial coefficients, 
$X_{t,\ell}=X_{t-1,\ell-1}+X_{t-1,\ell}$.
Therefore, $1-p=\frac{X_{t-1,\ell}}{X_{t,\ell}}$.
By using the recursive formula for the capacity $\widehat{\cC}_{t,\ell}$ described in Equation~(\ref{eq:capacityModelEIA:DIrecurssive}),
we are only left to prove that for all $2\leq \ell\leq t-1$,
$$
\log X_{t,\ell} = h(p)+p\log X_{t-1,\ell-1}+(1-p)\log X_{t-1,\ell}.
$$
This relation holds since
\begin{equation*}
	\begin{array}{l}
    h(p)+p\log X_{t-1,\ell-1}+(1-p)\log X_{t-1,\ell} \\
    =p\left(\log \left( \frac{X_{t,\ell}}{X_{t-1,\ell-1}}\right) +\log(X_{t-1,\ell-1})\right) 
  \\
  + (1-p)\left(\log \left( \frac{X_{t,\ell}}{X_{t-1,\ell}}\right) +\log(X_{t-1,\ell})\right)\\
    =p\log X_{t,\ell}+(1-p)\log(X_{t,\ell})\\
    =\log(X_{t,\ell}).
\end{array}
\end{equation*}
\end{IEEEproof}

\subsection{The EIA:DU Model - Constructions and Direct Part of Theorem~\ref{thm:capacityEIA}}\label{subsec:EIA:DU}
In this subsection, we study the EIA:DU model, that is, encoder informed all and decoder uninformed. Our main contribution is a construction of a capacity-achieving $\ell$-change $t$-write EIA:DU-ELM code for the zero-error case, 
which assures that 
$\cC_{t,\ell} \subseteq \cC_{t,\ell}^{EIA:DU,z}$.
That is, the direct part of Theorem~\ref{thm:capacityEIA} is proved.

Let us start with the first non-trivial case of $t=3$ and $\ell=2$.
Thus, we want to prove that $\cC_{3,2} \subseteq \cC_{3,2}^{EIA:DU,z} $.
Recall that, 
\begin{align*}
\cC_{3,2} \hspace{-0.5ex}=\hspace{-0.5ex}\Big\{(R_1,R_2,R_3)|& R_1\leq h(p_{1,0}), \\
& R_2\hspace{-0.5ex}\leq\hspace{-0.5ex} 1\hspace{-0.5ex}-\hspace{-0.5ex}p_{1,0}+p_{1,0} h(p_{2,1}),\\
& R_3\hspace{-0.5ex}\leq\hspace{-0.5ex} 1\hspace{-0.5ex}-\hspace{-0.5ex}p_{1,0} p_{2,1}, \text{ and } p_{1,0},p_{2,1}\hspace{-0.5ex}\in\hspace{-0.5ex}[0,0.5] \Big\}.
\end{align*}
which is achieved by setting $p_{3,0}=p_{3,1}=p_{2,0}=0.5$
in Equation~(\ref{eq:capacityModelEIA}). The next theorem states the existence of a construction of ELM codes for this case.

\begin{theorem}\label{thm.EIA:DU}
For any $\epsilon >0$ and $p_{1,0},p_{2,0},p_{2,1}\in [0,0.5]$,
there exists an explicit construction of a zero-error two-change three-write EIA:DU-ELM code satisfying $R_1 \geq h(p_{1,0})-\epsilon$, $R_2 \geq (1-p_{1,0}) h(p_{2,0})+p_{1,0}  h(p_{2,1})-\epsilon$, and $R_3 \geq (1-p_{1,0} p_{2,1})-\epsilon$. 
\end{theorem} 
Before presenting our construction for two-change three-write EIA:DU-ELM codes, we introduce the following family of WOM codes. We then use these WOM codes as component codes in our construction of EIA:DU-ELM codes.
Note that the WOM codes we use for our construction are given for $n\to \infty$, and thus our constructions for ELM codes use such $n$. 
\begin{definition}\label{defqary}
An $[n,2;M_1, M_2]_q^{EI:DU,z}$ two-write $q$-ary EI:DU WOM code for the zero-error case is a coding scheme comprising of $n$ $q$-ary cells.
It consists of two pairs of encoding and decoding maps $(\E_{q,1},\D_{q,1})$ and $(\E_{q,2},\D_{q,2})$ which are defined as follows:
\begin{enumerate}[(1)]
\item $\E_{q,1}: [M_1] \to [q]^n$ and $\D_{q,1}: Im(\E_{q,1}) \to [M_1]$ such that for all $m_1\in[M_1]$, $\D_{q,1}(\E_{q,1}(m_1))=m_1$. 
\item $\E_{q,2}: [M_2] \times Im(\E_{q,1}) \to [q]^n$ and $\D_{q,2}: Im(\E_{q,2}) \to [M_2]$ such that for all $(m_2,\bfc) \in [M_2] \times Im(\E_{q,1})$, $\E_{q,2}(m_2,\bfc) \geq \bfc$ and $\D_{q,2}(\E_{q,2}(m_2,\bfc))=m_2$.
\end{enumerate}
\end{definition}
We say that $\bfp=(p_0,p_1,\ldots,p_{m-1})$ is a \emph{probability vector} if $\sum_{i=0}^{m-1} p_i=1$ and $p_i \geq 0$ for all $i \in [m]$. 
We distinguish between an error-probability vector that is used in Definition~\ref{def:ELM_code}, and a probability vector.  An error-probability vector is a vector of error-probabilities, and not a probability vector, i.e., the sum of the elements of an error-probability vector does not need be 1.
For two positive integers $n,q$ and a probability vector $\bfp=(p_0,p_1,\ldots,p_{q-1})$, we denote by $\B(n,\bfp)$ the set of all length-$n$ $q$-ary vectors of constant composition $\bfw=(w_0,\ldots,w_{q-1})$, where $w_i=p_i n$ for $i\in [q]$
\footnote{We assume here that $p_i$ is a rational number and $n$ is large enough such that $p_i n$ is an integer for $i\in [q]$.}. 
Let $p_{j,i \rightarrow k}$ be the probability that on the $j$-th write, a cell in state $i$ is programmed to state $k$, $k\geq i$.

A family of two-write $q$-ary capacity-achieving EI:DU WOM codes was constructed recently by Shpilka~\cite{Sh14}. Particularly, given $\epsilon >0$ and probability vectors
$\bfp_{1,0},\bfp_{2,0},\ldots,\bfp_{2,q-2}$, Shpilka~\cite{Sh14} constructed a family of two-write $q$-ary EI:DU WOM codes that match these probability vectors on the first and second writes. We state this result formally.

\begin{lemma}\label{lem.WOM.qary}\cite{Sh14} 
For all $(j,i)\in \{(1,0),(2,0),(2,1),\ldots, (2,q-2)\}$, let $\bfp_{j,i}=(p_{j,i\rightarrow i},p_{j,i\rightarrow i+1},\ldots, p_{j,i\rightarrow q-1})$  be a probability vector. 
Then, for all $\epsilon>0$ there exists an $[n,2;M_1, M_2]_q^{EI:DU,z}$ two-write $q$-ary EI:DU WOM code satisfying:
\begin{enumerate}[\hspace{-0.05cm}{\tiny \textbullet}\hspace{-0.07cm}]
\item $Im(\E_{q,1}) 
\subseteq  \B(n,\bfp_{1,0})$ and $R_1=\frac{\log M_1}{n} \geq h(\bfp_{1,0})-\epsilon.$
\item For all $\bfc_{1} \in Im(\E_{q,1})$, $m_2\in[M_2]$, and $\bfc_2=\E_{q,2}(m_2,\bfc_{1})$, the following condition holds. 
For $i\in [q]$, let $\bfc^i_{2}$ be a length-$w_{1,i}$, $w_{1,i}=n p_{1,0\rightarrow i}$, substring of $\bfc_2$ at all locations $k$ with value $i$ before the second write, that is, $\bfc_{1,k}=i$. Then, $\bfc^i_2 \in \B(w_{1,i},\bfp_{2,i})$. Furthermore, $R_2 = \frac{\log M_2}{n} \geq \sum_{i=0}^{q-2} 
p_{1,0 \rightarrow i}  h(\bfp_{2,i})-\epsilon.$
\end{enumerate}
\end{lemma}

We refer to the family of WOM codes from Lemma~\ref{lem.WOM.qary} as  
an $[n,2;M_1, M_2]_q^{EI:DU}(\epsilon,\bfp_{1,0},\bfp_{2,0},\ldots,\bfp_{2,q-2})$ WOM code, where $M_1 = 2^{R_1n}$ and $M_2=2^{R_2n}$ are determined as the maximal possible values based on $\epsilon$, which tends to zero, and the probability vectors $\bfp_{j,i}$.

For the case $q=2$, for shorthand, given $p_{1,0\rightarrow 1}=p$ we denote these codes by $[n,2;M_1,M_2]^{EI:DU,z}(\epsilon, p)$ (where $p_{2,0\rightarrow 1}=0.5$).

Furthermore, using cooling codes, the work in \cite{CEKV18} provides the following family of binary WOM codes.
\begin{lemma}\label{lem.WOM.binary}
For all $p\in[0,0.5]$ and $\epsilon > 0$,
there exists a two-write binary WOM code $[n,2;M_1, M_2]^{EI:DU,z}(\epsilon,p)$ such that $M_1= \sum_{i=0}^{\tau} {n \choose i}$ and $M_2=2^{n-\tau-1}$, where $\tau= p n$. Therefore, for any $\epsilon >0$, there exists $n$ such that $R_1=\frac{\log M_1}{n} \geq h(p)-\epsilon$ and $R_2=\frac{\log M_2}{n} \geq 1-p-\epsilon$. 
\end{lemma}

We are now ready to present a construction of two-change three-write EIA:DU-ELM codes which establishes the result in Theorem~\ref{thm.EIA:DU}.
\begin{construction}\label{constructEIA:DU32}
Given $p_{1,0},p_{2,0},p_{2,1}\in [0,0.5]$ and $\epsilon>0$,
we construct an
$[n,3,2;M_1,M_2,M_3]^{EIA:DU,z}$ 
ELM code 
where $M_j=2^{nR_j}$ for $j =1,2,3$ such that $R_1 \geq h(p_{1,0})-\epsilon$, $R_2 \geq (1-p_{1,0}) h(p_{2,0})+p_{1,0} h(p_{2,1})-\epsilon$, and $R_3 \geq (1-p_{1,0} p_{2,1})-\epsilon$. 
We use the following two WOM codes.
\begin{enumerate}
\item Let $\bfp_{1,0}=(p_{1,0 \rightarrow 0}, p_{1,0 \rightarrow 1}, p_{1,0 \rightarrow 2})=( 1-p_{1,0},p_{1,0},0)$, $\bfp_{2,0}= (p_{2,0 \rightarrow 0}, p_{2,0 \rightarrow 1}, p_{2,0 \rightarrow 2}) =(0, p_{2,0}, 1- p_{2,0})$, and $\bfp_{2,1}= (p_{2,1 \rightarrow 1}, p_{2, 1\rightarrow 2}) =(1-p_{2,1},p_{2,1})$.
Let $C_1$ be an $[n,2;M_1, M_2]_3^{EI:DU,z}(\epsilon, \bfp_{1,0},\bfp_{2,0},\bfp_{2,1})$ two-write ternary EI:DU WOM code from Lemma \ref{lem.WOM.qary} with two pairs of encoder/decoder $(\E_{3,1},\D_{3,1})$ and $(\E_{3,2},\D_{3,2})$. 
\item Let $\rho_1=p_{1,0}  p_{2,1}$, and $C_2$ be an $[n,2;M_1', M_3]^{EI:DU,z}(\epsilon,\rho_1)$ two-write binary EI:DU WOM code from Lemma \ref{lem.WOM.binary} with two pairs of encoder/decoder $(\E_{2,1},\D_{2,1})$ and $(\E_{2,2},\D_{2,2})$. 
\end{enumerate}

The three pairs of encoder/decoder mappings $(\E^{EIA:DU}_j,\D^{EIA:DU}_j)$ for $j=1,2,3$ are defined as follows.  
\begin{enumerate}[\hspace{-0.2cm}]

\item \textbf{First write:}
$\E^{EIA:DU}_1(m_1)=\E_{3,1}(m_1)$ for all $m_1 \in [M_1]$. Similarly, 
$\D^{EIA:DU}_1(\bfc_1) =\D_{3,1}(\bfc_1)$.
Note that since we chose the probability to program level 2 in the first write of $C_1$ to be zero, the output of the encoder $\E_{3,1}$ is indeed a binary vector, so $\E^{EIA:DU}_1$ and $\D^{EIA:DU}_1$ are well defined.
\item \textbf{Second write:}
The idea is to use the second write encoder $\E_{3,2}$ of $C_1$ with the probability vectors $\bfp_{2,0}$ and $\bfp_{2,1}$, and notice that here we write all cells to levels 1 or 2. Then, we can view this ``ternary word'' as a binary word. Let $\bfc_1=(c_{1,1},\ldots,c_{1,n})\in Im(\E^{EIA:DU}_1)$ be the cell-state vector after the first write, and note that this is a binary vector.
The encoder/decoder $(\E^{EIA:DU}_2,\D^{EIA:DU}_2)$ are defined formally as follows.
For all $(m_2,\bfc_1) \in [M_2] \times Im(\E^{EIA:DU}_1)$, 
$$\bfc_2 =  \E^{EIA:DU}_2(m_2,\bfc_1)= \bfc_2'(\bmod 2),$$
where $\bfc_2' = \E_{3,2}(m_2,\bfc_1) \in [3]^n$.
Furthermore, for all $\bfc_2 \in Im(\E^{EIA:DU}_2)$, 
$$\D^{EIA:DU}_2(\bfc_2)=\D_{3,2}(\bfc'_2)=m_2,
$$ 
where $\bfc_2' = 2\cdot \textbf{1} - \bfc_2$, that is, $c'_{2,i}=1$ if $c_{2,i}=1$ and $c'_{2,i}=2$ if $c_{2,i}=0$.
\item \textbf{Third write:}
Let $\bfc_2$ be the cell-state vector after the second write. We note that the encoder on the third write knows the program-count vector 
$\bfv_2\in [3]^n$,
but the decoder does not have this information. Among the $n$ cells, there are $\rho_1  n$ cells which have been programmed twice, where $\rho_1 =p_{1,0} p_{2,1}$, and therefore (only) these cells cannot be programmed on this write. 
Hence, the encoder can interpret the vector 
$\bfv_2$ as a length-$n$ binary vector indicating for each cell whether it can be programmed on this write. We denote this vector by $\bfc''_2$, so 
$\bfc''_{2,i}=1$ 
if and only if $\bfv_{2,i}=2$. We will use the code $C_2$ to encode and decode on this write,
and we denote by $\bar{\bfc}$ the bitwise complement of a binary vector $\bfc$.
Specifically, 
the encoder/decoder mappings are defined as follows.
For all $m_3 \in [M_3]$ and $\bfv_2\in N_2$,  
$$\E_3^{EIA:DU}(m_3,\bfv_2)= \overline{\E_{2,2}(m_3,\bfc''_2)}.$$
Furthermore, 
for all $\bfc_3\in Im(\E_3^{EIA:DU})$,
$$\D_3^{EIA:DU} (\bfc_3)= \D_{2,2}(\overline{\bfc_3}).$$ 
\end{enumerate}
\end{construction}
To illustrate Construction \ref{constructEIA:DU32}, we present the following example.
\begin{example}
Let $n=7, p_{1,0}=3/7, p_{2,0}=1/2,$ and $p_{2,1}=1/3$, we construct a $[7,3,2;M_1,M_2,M_3]^{EIA:DU,z}$ three-change two-write ELM code as follows.
In the first write, we encode a message $m_1$ to obtain a binary vector of length 7, e.g., $\bfc_1=(1,1,1,0,0,0,0)$.
In the second write, to encode a message $m_2$, in the first step, we use the second write encoder $\E_{3,2}$ of the ternary code $C_1$ in Lemma \ref{lem.WOM.qary} with probability $\bfp_{2,0}=(0,1/2,1/2)$ and $\bfp_{2,1}=(2/3,1/3)$ to obtain  $\bfc_2'=\E_{3,2}(\bfc_1,m_2)$, e.g., $\bfc_2'=( 2,1,1,1,1,2,2)$. In the second step, we can replace symbol 2 by symbol 0 in the vector $\bfc_2'$ to obtain the binary vector $\bfc_2=(0,1,1,1,1,0,0)$. So, $\bfc_2$ is the output in the second write. We observe that it is not difficult to decode the vector $\bfc_2$ to obtain the message $m_2$. In the last write, the encoder has all information and know that the first cell is programmed twice and the program-count vector is $\bfv_2=(2,1,1,1,1,0,0)$. The encoder also view the vector $\bfv_2$ as a binary vector $\bfc_2''=(1,0,0,0,0,0,0)$ where the first cell (=1) is not programmable. Using the second-write encoder $\E_{2,2}$ of the EIU:DI WOM code $C_2$, we can encode the message $m_3$ to obtain $\bfc_3'=\E_{2,2}(m_3,\bfc_2'')$, e.g., $\bfc_3'=(1,0,0,0,1,1,1)$. We now take the bitwise complement of $\bfc_3'$ to obtain $\E_3^{EIA:DU}(m_3,\bfv_2)=\bfc_3=(0,1,1,1,0,0,0)$. So, all the three messages are $\bfc_1=(1,1,1,0,0,0,0), \bfc_2=(0,1,1,1,1,0,0),$ and $\bfc_3=(0,1,1,1,0,0,0)$.
\end{example}
We now present the proof of Theorem~\ref{thm.EIA:DU}.
\begin{IEEEproof}[Proof of Theorem 5]
Let $R_j(C_i)$ be the rate of the WOM code $C_i$ on the $j$-th write.
For any $\epsilon >0$ and $p_{1,0},p_{2,0},p_{2,1}\in [0,0.5]$, we choose the codes $C_1$ and $C_2$ in Construction~\ref{constructEIA:DU32} to satisfy
\[R_1(C_1) 
\geq h(\bfp_{1,0})-\epsilon= h(p_{1,0})-\epsilon,\]
where $\bfp_{1,0}=(1-p_{1,0},p_{1,0},0)$. 
\begin{align*}
R_2(C_1) & \geq  p_{1,0 \rightarrow 0}  h(\bfp_{2,0}) + p_{1,0 \rightarrow 1}  h(\bfp_{2,1})-\epsilon\\
&\geq (1-p_{1,0}) h(p_{2,0}) + p_{1,0} h(p_{2,1})-\epsilon,
\end{align*} 
and 
\[R_2(C_2)\geq 1-\rho_1-\epsilon=1-p_{1,0} p_{2,1}-\epsilon.\]
The result follows from the fact that the rate tuple of the two-change three-write ELM code is $(R_1(C_1),R_2(C_1),R_2(C_2))$. 
\end{IEEEproof}

The solution for the case $t=3,\ell=2$ is generalized for any $t$ and $\ell$ in the following theorem. 
\begin{theorem}\label{thm.EIA:DU.tl}
For all $t$ and $\ell,$
$\cC_{t,\ell}\subseteq \cC^{EIA:DU,z}_{t,\ell}$,
that is,
for any $\epsilon > 0$ and a rate $t$-tuple $(R_1,\ldots, R_t) \in
\cC_{t,\ell}$, 
there exists a zero-error $\ell$-change $t$-write EIA:DU ELM code $C$ such that its rate on the $j$-th write is at least $R_j - \epsilon$ for all $1 \leq j\leq t$, that is, $R_j(C) \geq R_j - \epsilon$. 
\end{theorem}

To prove Theorem \ref{thm.EIA:DU.tl}, we construct a zero-error $\ell$-change $t$-write ELM code. The idea is to generalize Construction \ref{constructEIA:DU32}. Hence, for any given $j$, we use $q$-ary EI:DU WOM code from Lemma \ref{lem.WOM.qary} to program all cells up to the two highest levels $q-1$ and $q-2$. So, the decoder can look at $q-1$ as 0 and $q-2$ as 1 to decode the original message. We now present the construction formally as follows.
\begin{construction}\label{const.EIA:DU.tl}
Given $p_{j,i} \in [0,0.5]$, for all $i \in [\ell+1]$, $1 \leq j \leq t$, and $\epsilon>0$, 
we construct an $[n,t,\ell; M_1,\ldots,M_t]^{EIA:DU,z}$ 
ELM code where $M_1 = {n \choose  p_{1,0}n }$
and $M_j=2^{nR_j}$ for $2\leq j \leq t$ such that $R_j \geq \sum_{i=0}^{m-1} Q_{j-1,i}h(p_{j,i}) -\epsilon$, where $Q_{j-1,i}$ is
defined in Equation~(\ref{eq:EIA:DIqProbFunc}). The $t$ pairs of encoder/decoder mappings $(\E_j^{EIA:DU},\D_j^{EIA:DU})$ 
are defined as follows.
\begin{enumerate}[\hspace{-0.2cm}]
\item \textbf{First write:}
Given $p_{1,0}$, we program all words of length $n$, weight $p_{1,0}n$ as on the first write of Construction \ref{constructEIA:DU32}.
Hence, $M_1= {n \choose p_{1,0}n}$
and the rate on the first write satisfies $R_1\geq h(p_{1,0}) -\epsilon$.
\item \textbf{$j$-th write, $2\leq j\leq t$:} 
Let $m=\min \{j,\ell\}$.
We denote the cell-state vector and the cell-program-count vector after the $j-1$ writes by $\bfc_{j-1}=(c_{j-1,1},\ldots,c_{j-1,n})\in Im(\E^{EIA:DU}_{j-1})$ and $\bfv_{j-1}=(v_{j-1,1},\ldots,v_{j-1,n}) \in \B(n,\bfq_{j-1}) \subset [m]^n$, respectively, where $\bfq_{j-1}=(Q_{j-1,0},Q_{j-1,1},\ldots,Q_{j-1,m-1})$ and $Q_{j,i}$ are defined in Equation~(\ref{eq:EIA:DIqProbFunc}).
To program on the $j$-th write, we use the two-write $(2m+1)$-ary WOM code from Lemma~\ref{lem.WOM.qary}, $[n,2;M_{1,j},M_{2,j}]_{2m+1}^{EI:DU,z}(\epsilon, \bfp_{1,0},\bfp_{2,0}, \bfp_{2,1},\ldots, \bfp_{2,2m-1})$ 
where $\bfp_{1,0}=(\bfq_{j-1},0,0,\ldots,0)$ 

and for all $i \in [m-1]$, 
$\bfp_{2,i}=(p_{2,i \rightarrow i},\ldots,p_{2,i \rightarrow 2m})=(0,\ldots,0,p_{j,i},1-p_{j,i})$
if $i$ is even and $\bfp_{2,i}=(p_{2,i \rightarrow i},\ldots,p_{2,i \rightarrow 2m})=(0,\ldots,0,1-p_{j,i},p_{j,i})$ 
if $i$ is odd. 
As in Lemma \ref{lem.WOM.qary}, $M_{2,j}=2^{R_j n}$ where $R_j \geq \sum_{i=0}^{m-1} Q_{j-1,i} h(\bfp_{2,i}) - \epsilon = \sum_{i=0}^{m-1} Q_{j-1,i} h(p_{j,i}) - \epsilon$ since $\bfp_{2,i}=(p_{2,i \rightarrow i},\ldots,p_{2,i \rightarrow 2m})=(0,\ldots,0,1-p_{j,i},p_{j,i})$ or $\bfp_{2,i}=(p_{2,i \rightarrow i},\ldots,p_{2,i \rightarrow 2m})=(0,\ldots,0,p_{j,i},1-p_{j,i})$.
Hence, on the $j$-th write, we choose $M_j = M_{2,j} = 2^{R_j n}$.
We denote the two pairs of the encoder/decoder of the used WOM code by $(\E_{m,1},\D_{m,1})$ and $(\E_{m,2},\D_{m,2})$.
The idea is to push all cells to the two highest levels and view the obtained word as a binary word. Hence, to decode correctly, the decoder only needs to know the cell-state vector after the $j$-th write which is a binary word.
We now define the encoder/decoder $(\E_j^{EIA:DU},\D_j^{EIA:DU})$ formally as follows.
For all each $m_j \in [M_j]$ and $\bfv_{j-1} \in Im(\E_{m,1})$
$$\bfc_j =  \E^{EIA:DU}_{j}(m_j,\bfv_{j-1})= \bfc_j'(\bmod 2),$$
where $\bfc_j' = \E_{m,2}(m_j,\bfv_{j-1}) \in [2m+1]^n$.
Furthermore, for all $\bfc_j \in Im(\E^{EIA:DU}_j)$, 
$$\D^{EIA:DU}_j(\bfc_j)=\D_{m,2}(\bfc'_j)=m_j,
$$ 
where $c'_{j,i}=2m-1$ if $c_{j,i}=1$ and $c'_{j,i}=2m$ if $c_{j,i}=0$.
\end{enumerate}
\end{construction}

\begin{IEEEproof}[Proof of Theorem \ref{thm.EIA:DU.tl}]
Given all parameters as in Construction \ref{const.EIA:DU.tl}, the rate of this ELM code on the first write is $R_1\geq h(p_{1,0})-\epsilon$. Now, we consider the $j$-th write. Since we used the WOM code in Lemma \ref{lem.WOM.qary} to program the $j$-th write of the ELM code, the rate on this write is exactly the rate on the second write of the used WOM code. Hence, the rate in the $j$-th write of the ELM code is $R_j \geq \sum_{i=0}^{m-1} Q_{j-1,i}h(p_{j,i})- \epsilon$.
\end{IEEEproof}
\begin{remark}
In this section, we provide an explicit construction of zero-error two-change three-write EIA:DU ELM code and generalize the result to construct a zero-error $\ell$-change $t$-write EIA:DU ELM code. Since Shpilka \cite{Sh14} provided a pair of polynomial time encoding/decoding algorithms of a family of two-write WOM codes, the encoder and decoder in Theorem \ref{thm.EIA:DU.tl} also run in polynomial time. As shown in Theorem \ref{thm.EIA:DU.tl}, using these constructions, we can achieve any rate in the capacity region and thus achieve the maximum sum-rate when the length $n$ tends to infinity. However, for a fixed value of $n$, we can only achieve a high sum-rate but can not achieve the maximum sum-rate. Furthermore, Shpilka's technique only works for large block length\cite{Sh14}. Hence, for small value of block length $n$, we need other constructions to obtain a high sum-rate, for example, Construction \ref{constructEIPDUtl} that will be presented later.
\end{remark}

\subsection{The EIA:DIA Model - Converse Part of Theorem~\ref{thm:capacityEIA}}\label{subsec:EIA:DIA}
In this section, we prove the converse part of Theorem~\ref{thm:capacityEIA} for the EIA:DIA model $\epsilon$-error case. 
That is, we prove that $\cC_{t,\ell}^{EIA:DIA,\epsilon}\subseteq \cC_{t,\ell}$.

For this direction we need to prove
that if there exists an $[n,t,\ell;M_1,\ldots, M_t]^{EIA:DIA,\bfp_{e}}$ ELM code where $\bfp_{e}=(p_{e_1},\ldots,p_{e_t})$, then 
$$\left(\frac{\log M_1}{n}-\epsilon_1, \frac{\log M_2}{n}-\epsilon_2, \ldots, \frac{\log M_t}{n}-\epsilon_t\right)\in {\cC}_{t,\ell},$$
where $(\epsilon_1,\epsilon_2,\ldots,\epsilon_t)$ tends to $\bf0$ if
$\bfp_{e}$ tends to $\bf0$ and $n$ tends to infinity. 
In our proof $\epsilon_j=\frac{H(p_{e_j})+p_{e_j}\log (M_j)}{n}$,
and therefore $\epsilon_j\to 0$ when $p_{e_j}\to 0$ and $n\to \infty$.

Let $X_j$ be a length-$n$ binary vector where $X_{j,k}=1$ if and only if the $k$-th cell is intended to be programmed on the $j$-th write.
Similarly, $Y_j$, is a length-$n$ binary vector, where $Y_{j,k}=1$ 
if and only if the value of the $k$-th cell was successfully changed on the $j$-th write, that is, $Y_j=\bfc_j\oplus\bfc_{j-1}$.
Note that the encoder knows the number of times each cell was programmed. Therefore, we can assume that a cell is not intended to be programmed more than $\ell$ times.
Furthermore, the decoder also knows the number of times each cell was programmed. Thus we assume that $X_j=Y_j$ where $X_j$ is the encoded word and $Y_j$ is the input of the decoder. 

Let $S_1, \ldots, S_t$ be independent random variables, where $S_j$ is uniformly distributed over the messages set $[M_j]$,
and $\hat{S}_j$ is the decoding result on the $j$-th write.
Let $V_{j}$ be an independent random variable on $N_j$, the set of all cell-programs-count vectors after the first $j$ writes. 
The data processing yields the following Markov chain:
    $$
	S_j|V_{j-1} \text{ --- } X_j|V_{j-1} \text{ --- } Y_j|V_{j-1} \text{ --- } \hat{S}_j|V_{j-1}
	$$
	and	therefore, $I(X_j;Y_j|V_{j-1}) \ge I(S_j;\hat{S}_j|V_{j-1}).$
	
    Additionally,
	\begin{equation*}
	\begin{array}{ll}
	I(S_j;\hat{S}_j|V_{j-1})& = H(S_j|V_{j-1})-H(S_i|\hat{S}_j,V_{j-1})\\
	& \ge H(S_j) - H(S_j | \hat{S}_j) \\
	& \ge \log (M_j)-H(p_{e_j})-p_{e_j}\log (M_j).
	\end{array}
	\end{equation*}
	The first inequality follows from the independence of $V_{j-1}$ and $S_{j}$ which implies that $H(S_j|V_{j-1})=H(S_j)$, and from the fact that conditioning does not increase the entropy. 
    The second inequality follows from Fano's inequality \cite[p.~38]{CT91} $H(S_j|\hat{S}_j)\le H(p_{e_j})+p_{e_j}\log (M_j)$.

Let $L$ be an index random variable, which is uniformly distributed over the index set $[n]$. Since $L$ is independent of all other random variables we get
	\begin{equation*}
	\begin{array}{ll}
	\hspace{-2ex}\dfrac{1}{n}I(X_j;Y_j|V_{j-1}) & \hspace{-1.5ex} \le \dfrac{1}{n}H(Y_j|V_{j-1})\\
	&\hspace{-1.5ex} \overset{(a)}{\le}  \dfrac{1}{n} \sum_{k=0}^{n-1} H(Y_{j,k}|V_{j-1,k})\\
	&\hspace{-1.5ex} \overset{(b)}{=} H(Y_{j,L}|V_{j-1,L},L)\\
	&\hspace{-1.5ex} \overset{(c)}{\le}  H(Y_{j,L}|V_{j-1,L})\\
	&\hspace{-1.5ex} = \sum_{i=0}^{\ell} Pr(V_{j-1,L}\hspace{-0.75ex}=i)H(Y_{j,L}|V_{j-1,L}=i)\\
	&\hspace{-1.5ex} \overset{(d)}{=} \sum_{i=0}^{\ell-1} Pr(V_{j-1,L}\hspace{-0.75ex}=i)H(Y_{j,L}|V_{j-1,L}=i),
	\end{array}
	\end{equation*}
	where steps $(a)$ and $(c)$ follow from the fact that entropy of a vector is not greater than the sum of the entropies of its components, and conditioning does not increase the entropy. Step $(b)$ follows from the fact that 	
	\begin{equation*}
	\begin{array}{ll}
	\hspace{-1.5ex}H(Y_{j,L}|V_{j-1,L},L)&\hspace{-1.5ex}=\sum_{k=0}^{n-1}Pr(L=k)H(Y_{j,k}|V_{j-1,L},L=k)\\
	&\hspace{-1.5ex}=\dfrac{1}{n} \sum_{k=0}^{n-1}H(Y_{j,k}|V_{j-1,k}),
	\end{array}
	\end{equation*} 
	and step $(d)$ follows from $H(Y_{j,L}|V_{j-1,L}=\ell)=0$.

Now, we set 
	$p_{j,i}= Pr(X_{j,L}=1|N_{j-1,L}=i)= Pr(Y_{j,L}=1|N_{j-1,L}=i)$,
	and thus we can conclude that 
    $Q_{j,i}=Pr(N_{j,L}=i)$
    where $Q_{j,i}$ is calculated in Equation~(\ref{eq:EIA:DIqProbFunc}),
	and then
    \begin{equation*}
	\begin{array}{ll}
	\hspace{-1ex}
	\dfrac{\log (M_j)}{n}-\epsilon_j	& \le 
	\dfrac{1}{n}I(X_j;Y_j|N_{j-1}) \\
	& \le  \sum_{i=0}^{\ell-1} Pr(N_{j-1,L}=i)H(Y_{j,L}|N_{j-1,L}=i)\\
	& =\sum_{i=0}^{\ell-1} Q_{j-1,i}h\left(p_{j,i}\right),
	\end{array}
    \end{equation*}
	where $\epsilon_j=\frac{H(p_{e_j})+p_{e_j}\log(M_j)}{n}$,
	and the converse part is implied.

By Theorem~\ref{thm.EIA:DU.tl} in Subsection~\ref{subsec:EIA:DU} 
and by the proof of the converse part in Subsection~\ref{subsec:EIA:DIA}
we completed the proof of Theorem~\ref{thm:capacityEIA}.
Furthermore by Theorem~\ref{thm:EIA_maxSumRate}
we conclude the following corollary.
\begin{corollary}
For all $t$ and $\ell$,
$\cC_{t,\ell}=\widehat{\cC}_{t,\ell}$ is the capacity region for all the EIA models for both the zero-error and the $\epsilon$-error cases and is denoted by $\cC_{t,\ell}^{EIA}$. The maximum sum-rate of all the EIA models is $\cR^{EIA}_{t,\ell}= \log \sum_{i=0}^{\ell} {t \choose i}$.
\end{corollary}

\section{The Capacity of the EIP:DIA Model}\label{sec:EIP:DIA} 
In this section we discuss the capacity region and the maximum sum-rate of the EIP:DIA model.
Recall that if $\ell=1$ then by definition, EIP is equivalent to EIA and this model is equivalent to the known WOM model. Thus, in this section we assume that $\ell>1$.
We focus on the $\epsilon$-error case and present the capacity region of this model.
The zero-error case is harder to solve, and is left for future research. However the $\epsilon$-error case provides an upper bound for the zero-error case.
Note that 
the EU:DI WOM model is simpler than the EIP:DIA ELM model, and even though its exact capacity for the zero-error case is still not known for general $t$.

As done in the EIA models, let us denote by $\bfc_j$, $j\in[t+1]$, the length-$n$ binary vector which represents the memory state after the $j$-th write, where $\bfc_0=\bf0$. 

For $1\leq j\leq t$ and $i\in[\ell+1]$, we define the probabilities $p_{j,0}$, $p_{j,1}$, and $Q_{j,i}$ as follows.
$p_{j,k}$ is the probability of programming a cell on the $j$-th write given that the value of this cell was $k$, $k\in \{0,1\}$,
and $Q_{j,i}$ is the probability of a cell to be programmed exactly $i$ times after the first $j$ writes.
Additionally, let $Q_{j,e}, Q_{j,o}$ be the probability of a cell to be programmed an even, odd number of times after the first $j$ writes, respectively. 
Formally, $Q_{j,i}$ 
is defined recursively by using the probabilities $p_{j',0}$ and $p_{j',1}$ for $j'\leq j$. We now assume that $\ell$ is even. The case of an odd $\ell$ is defined similarly. We define $Q_{j,i}$ for $j>0$ as follows. For even $i\ge 0$,
\begin{equation}
\label{eq:EIP:DIAqProbFunc1}
\begin{array}{rl}
\hspace{-2.5ex}
Q_{j,i}& =\begin{cases}
Q_{j-1,i-1} p_{j,1}+Q_{j-1,i}(1-p_{j,0}), &  \text{if } 0< i< \ell,\\
Q_{j-1,i-1} p_{j,1}+Q_{j-1,i}, &  \text{if } i= \ell,\\
Q_{j-1,i}(1-p_{j,0}), &  \text{if } i=0,
\end{cases}
\end{array}
\end{equation}
and for odd $i>0$, $Q_{j,i}=Q_{j-1,i-1} p_{j,0}+Q_{j-1,i}(1-p_{j,1})$. The base $j=0$, is $Q_{0,0}=1$ and $Q_{0,i}=0$ for $i>0$. Furthermore, let $Q_{j,e}=\sum_{i=0}^{\ell/2}Q_{j,2i}$ and $Q_{j,o}=\sum_{i=1}^{\ell/2}Q_{j,2i-1}$.

Next, we define the rates region $\widetilde{\cC}_{t,\ell}$ which will be proved to be the capacity region of the EIP:DIA model for the $\epsilon$-error case. 
We present here the definition for even $\ell$, while the odd case is defined similarly.
\begin{equation}
\label{eq:capacityModelEIP:DIA2}
\begin{array}{ll}
\hspace{-1.1ex} \widetilde{\cC}_{t,\ell}\hspace{-0.5ex} =\hspace{-0.5ex}  
\Big\{ \hspace{-0.5ex} (R_1,R_2,\ldots, 
R_t)  | 
& \hspace{-2ex} \forall 1\leq j\leq t: \\
& \hspace{-18ex} 
R_j \leq Q_{j-1,o}h(p_{j,1})
 +(Q_{j-1,e}-Q_{j-1,\ell})h(p_{j,0}),\\
& \hspace{-18ex} p_{j,0}\hspace{-0.2ex},\hspace{-0.2ex}p_{j,1}\hspace{-0.5ex}\in\hspace{-0.5ex} [0,0.5]
\text{ and} \ Q_{j,e}, Q_{j,o},Q_{j,\ell} \text{ are defined above} \Big\}.\\
\end{array}
\ignorespacesafterend
\noindent
\end{equation}

For example, for $t=3,\ell=2$, we have that
\begin{align*}
\widetilde{\cC}_{3,2}
=\cC_{3,2} \hspace{-0.5ex}=\hspace{-0.5ex}\Big\{&(R_1,R_2,R_3)| R_1\leq h(p_{1,0}), \\
&\hspace{1ex}R_2\hspace{-0.5ex}\leq\hspace{-0.5ex} 1\hspace{-0.5ex}-\hspace{-0.5ex}p_{1,0}\hspace{-0.5ex}+\hspace{-0.5ex}p_{1,0}h(p_{2,1}),\\
&\hspace{1ex}R_3\hspace{-0.5ex}\leq\hspace{-0.5ex} 1\hspace{-0.5ex}-\hspace{-0.5ex}p_{1,0}p_{2,1}, \text{and } p_{1,0},p_{2,1}\hspace{-0.5ex}\in\hspace{-0.5ex}[0,0.5] \Big\},
\end{align*}
which is achieved by substituting $p_{3,0}=p_{3,1}=p_{2,0}=0.5$
in Equations
(\ref{eq:capacityModelEIA})  
and (\ref{eq:capacityModelEIP:DIA2}). 
Using the region $\widetilde{\cC}_{t,\ell}$, the next theorem characterizes the capacity region of the EIP models for the $\epsilon$-error case.

\begin{theorem}
	\label{thm:capacityEIP:DIA}
	The rates region $\widetilde{\cC}_{t,\ell}$ is the capacity region of $t$-write $\ell$-change ELM EIP:DIA model for the $\epsilon$-error case. 
   That is, $\widetilde{\cC}_{t,\ell}=\cC_{t,\ell}^{EIP:DIA,\epsilon}$.
\end{theorem}
\begin{IEEEproof}
To show the achievable region, we should prove
that for each $\epsilon>0$ and $(R_1, R_2, \ldots, R_t)\in \widetilde{\cC}_{t,\ell}$, there exists an \\ $[n,t;M_1,\ldots, M_t]_{t,\ell}^{EIP:DIA, \bfp_e}$ ELM code,
where for all $1 \leq j \leq t$, $\frac{\log M_j}{n}\ge R_j-\epsilon$ and
$\bfp_e=(p_{e_1},\ldots,p_{e_t}) \leq (\epsilon,\ldots,\epsilon)$.
We use the well-known random channel-coding theorem \cite[p.~200]{CT91} on each write.
We describe the encoding and decoding on each write.

The $j$-th write presents a DMC with the input length-$n$ binary vector $X_j$ and the output is $(Z_{j-1},Y_j)$, where $Z_{j-1}\in [\ell+1]^n$ represents the times each cell was programmed before the $j$-th write, and $Y_j\in [2]^n$ represent the state of the memory after the $j$-th write.
Let $x_j=X_{j,k}$, $z_j=Z_{j-1,k}$, and $y_j=Y_{j,k}$ for some index $k$.
By the random coding theorem, for $n$ large enough, the following region is achievable
\begin{equation*}
\begin{array}{ll}
\hspace{-1ex} \Big\{ \hspace{-0.5ex} (R_1,\ldots,R_t)  | & \forall 1\leq j\leq t, R_j \leq I(x_j;y_j)\Big\}.
\end{array}
\end{equation*}

By the definitions and notations of the probabilities $p_{j',i'}$ and $Q_{j',i'}$,
\begin{align*}
    I(x_j;(z_{j-1},y_j))
    &{=}H(z_{j-1},y_{j})-H(z_{j-1},y_{j} \vert x_{j}) \\
    &=H(z_{j-1}) +H(y_{j}|z_{j-1})-H(z_{j-1},y_{j} \vert x_{j}) \\
    &\overset{(a)}{=}H(z_{j-1}) +H(y_{j}|z_{j-1})-H(z_{j-1}) \\
    &=H(y_{j}|z_{j-1})\\
    &=\sum_{i=0}^{\ell} Pr(z_{j-1}=i)H(y_{j}|z_{j-1}=i)\\
    &\overset{(b)}{=}\sum_{i=0}^{\ell-1} Pr(z_{j-1}=i)H(y_{j}|z_{j-1}=i)\\
    &=\sum_{i=1}^{\ell/2} \left(Q_{j-1,2i-1}h\left(p_{j,1}\right)+
 Q_{j-1,2i-2}h\left(p_{j,0}\right)
	\right)\\
    &=Q_{j-1,o}h(p_{j,1})+(Q_{j-1,e}-Q_{j-1,\ell})h(p_{j,0}).
\end{align*}
Step $(a)$ follows from $H((z_{j-1},y_j)|x_i)=H(z_{j-1}|x_j)$ since $y_j$ is a function of $x_j,z_{j-1}$, and $H(z_{j-1}|x_j)=H(z_{j-1})$ because $z_{j-1}$ is independent on $x_j$. Step $(b)$ is implied by $H(y_j|z_{j-1}=\ell)=0$.

Hence, we can achieve the region $\widetilde{\cC}_{t,\ell}$ for the $\ell$-change $t$-write W$\ell$M EIP:DIA model for the $\epsilon$-error case.

The proof of the converse part, $\cC_{t,\ell}^{EIP:DIA,\epsilon}\subseteq \widetilde{\cC}_{t,\ell}$, is similar to the proof of this part in Theorem~\ref{thm:capacityEIA}, and hence is deferred to Appendix~\ref{app:proofs}.
\end{IEEEproof}

We could also present a family of capacity achieving codes using the binary erasure channel (BEC).
Note that on the $j$-th write, both the encoder and the decoder know $\bfc_{j-1}$, the state of the memory before writing the new data,
while the decoder also knows $\bfv_{j-1}$, the number of times each cell was programmed before the $j$-th write.
Therefore, the encoder on the $j$-th write treats the one and the zero cells separately.
On the cells with value one, the encoder 
writes zero with probability $p_{j,1}$
(for example by using a constant weight code),
while for the zero cells,  
the decoder knows which cells have been already programmed $\ell$ times before the $j$ write. Thus, the encoding on the zero cells can be represented as encoding over BEC with erasure probability $Q_{\ell}/Q_e$.
The capacity of the BEC with erasure probability $\pi$ and probability $\alpha$ for occurrence one in the encoded vector is $(1-\pi)h(\alpha)$~\cite[p.~188]{CT91}.
By substituting $p_{j,0}=\alpha$ and $\pi=Q_{\ell}/Q_e$,
we get the rate on the $j$-th write $ Q_{j-1,o}h(p_{j,1})
 +(Q_{j-1,e}-Q_{j-1,\ell})h(p_{j,0})$.

The following theorem is an immediate result deduced by the definitions of $\widetilde{\cC}_{t,\ell}$ and ${\cC}_{t,\ell}$ and Theorems~\ref{thm:capacityEIA} and~\ref{thm:capacityEIP:DIA}. 
\begin{theorem}
\label{thm:comparing_ell2}
    For $\ell=2$ the capacity region of the EIP:DIA model for the epsilon error case is equal to the capacity region for the EIA models, i.e., 
    $\cC_{t,2}^{EIP:DIA,\epsilon}=\cC_{t,2}^{EIA}$.
\end{theorem}

In Section~\ref{sec:comparison:EIAvsEIP:DIA}, we compare between the 
 EIP:DIA model which was discussed in this section,
and the EIA models, which were presented in Section~\ref{sec:EIA}.

\section{A Comparison between the EIA Models and the EIP:DIA Model}
\label{sec:comparison:EIAvsEIP:DIA}
In this section we compare between the EIA models and the EIP:DIA model.
The capacity of the EIA models,
$\cC^{EIA:DY,g}_{t,\ell}$ for $g\in \{z,\epsilon\}$ and $Y\in \{IA,IP,U\}$, was stated in Section~\ref{sec:EIA} to be equal to 
$\cC_{t,\ell}$,
while in Section~\ref{sec:EIP:DIA} we presented 
the capacity region of the EIP:DIA model for the $\epsilon$-error case, 
$\widetilde{\cC}_{t,\ell} =\cC^{EIP:DIA,\epsilon}_{t,\ell}$.

The next theorem proves that for $t>\ell\geq 3$ the maximum sum-rate of the EIP:DIA model for the epsilon-error case is smaller than the
maximum sum-rate of the EIA models.
Hence, the capacity region $\cC^{EIP:DIA,\epsilon}_{t,\ell}$ is a proper subset of the capacity region 
$\cC^{EIA}_{t,\ell}$ for these parameters.
Recall that for $\ell=2$ these regions were shown to be the same in
Theorem~\ref{thm:comparing_ell2}, and therefore the maximum sum-rates of these models for $\ell=2$ are the same too.

\begin{theorem}\label{thm:comparing1}
For $t>\ell\ge 3$,
$\cR^{EIP:DIA,\epsilon}_{t,\ell} <
\cR^{EIA}_{t,\ell}$,
and hence 
$\cC^{EIP:DIA,\epsilon}_{t,\ell} \subsetneq
\cC^{EIA}_{t,\ell}$.
\end{theorem}
\begin{IEEEproof}
Let $\widetilde{\bfR}=(\widetilde{R}_1,\widetilde{R}_2,\ldots, \widetilde{R}_t )$ be a rate tuple which achieves the maximum sum-rate $\cR^{EIP:DIA,\epsilon}_{t,\ell}$,
and we denote by ${\widetilde{p}}_{j,0}$, ${\widetilde{p}}_{j,1}$, and $\widetilde{Q}_{j,i}$, $1\le j\le t$ and $i\in[\ell+1]$, the probabilities which attain $\widetilde{\bfR}$ in $\widetilde{\cC}_{t,\ell}$.

Now we present a rate tuple ${\bfR}=({R}_1,{R}_2,\ldots, {R}_t)\in {\cC}_{t,\ell}> \widetilde{\bfR}$. Then, we conclude that $\bfR\in \cC^{EIA}_{t,\ell}\setminus \cC^{EIP:DIA,\epsilon}_{t,\ell}$,
which implies that $\cR^{EIP:DIA,\epsilon}_{t,\ell} < \cR^{EIA}_{t,\ell}$
and 
$\cC^{EIP:DIA,\epsilon}_{t,\ell} \subsetneq \cC^{EIA}_{t,\ell}$.

We assume now that $\ell$ is even, while the proof for the odd case is similar.
Since $\widetilde{\bfR}$ is maximal rate tuple we have $\widetilde{p}_{t-1,0}=\widetilde{p}_{t,0}=\widetilde{p}_{t,1}=0.5$.
For all $j$ and $i$, $1\le j \le t-2$ and $i\in [\ell]$, we define $p_{j,i}=\widetilde{p}_{j,i'}$ where $i'= i \mod 2$.
In addition, for all $i\in [\ell-1]$, $p_{t-1,i}=0.5$,  $p_{t-1,\ell-1}=\widetilde{p}_{t-1,1}$, and for all $i$, $p_{t,i}=0.5$.

Thus, for all $j$ and $i$, $1\le j \le t-2$ and $i\in [\ell]$, $R_j=\widetilde{R}_j$ and $Q_{j,i}=\widetilde{Q}_{j,i}$.
For the $(t-1)$-th write we have,
${R}_{t-1}= 1-\widetilde{Q}_{t-2,\ell-1}-\widetilde{Q}_{t-2,\ell}+
\widetilde{Q}_{t-2,\ell-1}h(\widetilde{p}_{t-1,1})$
while $\widetilde{R}_{t-1}= \widetilde{Q}_{t-2,o}h(\widetilde{p}_{t-1,1})+(\widetilde{Q}_{t-2,e}-\widetilde{Q}_{t-2,\ell})$, and for the last write
${R}_{t}= \widetilde{R}_{t}=1-\widetilde{Q}_{t-1,\ell}$.

Now we prove that $\widetilde{p}_{t-1,1}<0.5$ which immediately implies that ${R}_{t-1}>\widetilde{R}_{t-1}$ and thus completes the proof.
Recall that $\widetilde{R}_t=1-\widetilde{Q}_{t-1,\ell}=1-\widetilde{Q}_{t-2,\ell}-\widetilde{Q}_{t-2,\ell-1}\widetilde{p}_{t-1,1}$.
Thus, given the probabilities for the first $t-2$ writes, in order to achieve the maximal rate tuple $\widetilde{\bfR}$, we have to maximize $\widetilde{R}_{t-1}+\widetilde{R}_{t}$.
That is, we choose $\widetilde{p}_{t-1,1}$ which maximizes 
$\widetilde{Q}_{t-2,o}h(\widetilde{p}_{t-1,1})-
\widetilde{Q}_{t-2,\ell-1}\widetilde{p}_{t-1,1}$.
The derivative is 
$\widetilde{Q}_{t-2,0}\log ( \frac{1-\widetilde{p}_{t-1,1}}{\widetilde{p}_{t-1,1}})
-\widetilde{Q}_{t-2,\ell-1}$,
and the maximum is obtained for 
$\widetilde{p}_{t-1,1}=1 / (1+2^{{\widetilde{Q}_{t-2,\ell-1}}/{\widetilde{Q}_{t-2,o}}})$.
Since $\widetilde{\bfR}$ is maximal and $t>\ell\ge 3$, we have ${\widetilde{Q}}_{t-2,\ell-1}>0$, and therefore 
$\widetilde{p}_{t-1,1}\ne 0.5$.
\end{IEEEproof}

We can summarize the results regrading the capacity region of the EIP:DIA model
in the following corollary.
\begin{cor}
\label{cor:comparisonEIAvsDIPEIA}
For all $t > \ell$ the following holds
 $$
 \cC^{EIP:DIA,\epsilon}_{t,\ell} =
 \widetilde{\cC}_{t,\ell} \subseteq
 {\cC}_{t,\ell} =\cC^{EIA}_{t,\ell}.
$$
Furthermore, 
\begin{itemize}
\item
    For $t > \ell=2$ all these regions are equal, in particular,
    $\cC^{EIP:DIA,\epsilon}_{t,2}=\cC^{EIA}_{t,2}$.
\item
    For $t>\ell\ge 3$,  
   $\cC^{EIP:DIA,\epsilon}_{t,\ell}\subsetneq \cC^{EIA}_{t,\ell}$ and 
   $\cR^{EIP:DIA,\epsilon}_{t,\ell}< \cR^{EIA}_{t,\ell}$.
\end{itemize}
\end{cor}

\section{The Capacity of EU:DIA Model}\label{sec:EU:DIA} 

In this section we study the EU:DIA model for the $\epsilon$-error case,
and provide the capacity region of this model.
As in the EIP:DIA model, the capacity region for the zero-error case and the exact maximum sum-rate are left for future research.

For $1\leq j\leq t$ and $i\in[\ell+1]$,
let $p_{j}$ be the probability of programming a cell on the $j$-th write,
and $Q_{j,i}$ denotes the probability of a cell to be programmed exactly $i$ times on the first $j$ writes.
Additionally, let $Q_{j,e}, Q_{j,o}$ be the probabilities of a cell to be programmed even, odd number of times on the first $j$ writes, respectively.
Formally, $Q_{j,i}$ 
is defined recursively
by using $p_{j'}$ probabilities for $j' \le j$.
For $j\ge 1$,

\begin{equation}
\label{eq:EU:DIqProbFunc}
\begin{array}{rl}
\hspace{-2ex}Q_{j,i}&=\begin{cases}
Q_{j-1,i-1} p_{j}+Q_{j-1,i}(1-p_{j}), &  \text{if } 0< i\le \ell,\\
Q_{j-1,i-1} p_{j}+Q_{j-1,i}, &  \text{if } i=\ell, \\
Q_{j-1,i}(1-p_{j}), &  \text{if } i=0,
\end{cases}
\end{array}
\end{equation}
where $Q_{0,0}=1$ and $Q_{0,i}=0$ for $i>0$.

Then, we define the region $\overline{\cC}_{t,\ell}$
which is proved later in this section to be the capacity region $\cC_{t,\ell}^{EU:DIA,\epsilon}$.
\begin{equation}
\label{eq:capacityModelEIP:DI2}
\begin{array}{ll}
\hspace{-1ex} \overline{\cC}_{t,\ell}\hspace{-0.5ex} =\hspace{-0.5ex}  
\Big\{ \hspace{-0.5ex} (R_1,R_2,\ldots, 
R_t)  | 
& \hspace{-2ex} \forall 1\leq j\leq t: \\
& \hspace{-2ex} R_j \leq h(p_{j})-Q_{j-1,\ell}h(p_j) ,\\ 
& \hspace{-2ex}p_{j}\in [0,0.5],\ Q_{j,\ell} \text{ is defined above} \Big\}.\\
\end{array}
\ignorespacesafterend
\noindent
\end{equation}

The next theorems establish the capacity region of the EU:DIA model for the $\epsilon$-error case and compare between this model and the EIP:DIA model.
The techniques applied for the EU:DIA model are very similar to the proofs in Section~\ref{sec:EIP:DIA}.
The proofs of Theorems~\ref{thm:capacityEU:DIA} and~\ref{thm:comparing2} 
are similar to the proofs of Theorems~\ref{thm:capacityEIP:DIA} and~\ref{thm:comparing1}, respectively. 
Therefore, these proofs are moved to Appendix~\ref{app:proofs}.

\begin{theorem}\label{thm:capacityEU:DIA}
The rates region $\overline{\cC}_{t,\ell}$ is the capacity region of $t$-write $\ell$-change ELM EU:DIA model for the $\epsilon$-error case. That is,
$\overline{\cC}_{t,\ell}= \cC_{t,\ell}^{EU:DIA,\epsilon}$.
\end{theorem}

\begin{theorem}\label{thm:comparing2}
For $t>\ell\ge 2$,
$\cR^{EU:DIA,\epsilon}_{t,\ell} <
\cR^{EIP:DIA,\epsilon}_{t,\ell}$,
and hence 
$\cC^{EU:DIA,\epsilon}_{t,\ell} \subsetneq
\cC^{EIP:DIA, \epsilon}_{t,\ell}$.
\end{theorem}

\section{The EIP:DU Model}\label{sec:EIP:DU}
In this section, we study the EIP:DU model and its sum-rate. First, we note that $\cC^{EIP:DU,\epsilon}_{t,\ell}\subseteq\cC^{EIP:DIA,\epsilon}_{t,\ell}$ for all $t, \ell$,
and thus, 
\begin{equation}\label{eq1EIPDU}
    \cR^{EIP:DU,\epsilon}_{t,\ell} \leq \cR^{EIP:DIA,\epsilon}_{t,\ell} \le \log \sum_{i=0}^{\ell} {t \choose i}.
\end{equation}
That is, we obtain an upper bound of the maximum sum-rate $\cR^{EIP:DU,\epsilon}_{t,\ell}$. 
Note that for $t>\ell \ge 3$ this upper bound is not tight (Theorem~\ref{thm:comparing1}).
We are now interested in some good lower bounds for the maximum sum-rate.
Our goal is to provide several constructions with high sum-rate. 
We first present a general construction for the zero-error case and then show how to obtain higher sum-rate for the $\epsilon$-error case with $t=3,\ell=2$. 

The following construction provides a family of $\ell$-change $t$-write EIP:DU ELM codes for the zero-error case. 
\begin{construction}\label{constructEIPDUtl}
Let $(k_1,\dots,k_{\ell})$ be such that $1 \leq k_i \leq t$  for $1 \leq i \leq \ell$ and $\sum_{i=1}^{\ell} k_i =t.$ 
Let $[n,k_i;M_{j_i+1},\ldots,M_{j_i+k_i}]^{EI:DU,z}$ be a binary $k_i$-write EI:DU WOM code for $1 \leq i\leq \ell$ with sum-rate $R_i$ where $j_1=0$ and $j_i=\sum_{r=1}^{i-1}k_r$. Each of which consists of $n$ bits and $k_i$ pairs of encoding and decoding maps $(\E^{EI:DU}_{j_i+h},\D^{EI:DU}_{j_i+h})$ for $1 \leq h \leq k_i$. 
We define an $[n,t,\ell;M_1,\ldots,M_t]^{EIP:DU,z}$ $\ell$-change $t$-write ELM code
consists of $n$ bits and $t$ pairs of encoders and decoders $(\E^{EIP:DU}_{j},\D^{EIP:DU}_{j})$ where $\E^{EIP:DU}_j=\E^{EI:DU}_j$ and $\D^{EIP:DU}_j=\D^{EI:DU}_j$ for $1 \leq j \leq t$.
\end{construction}
The maximum sum-rate of the ELM codes from Construction~\ref{constructEIPDUtl} is $R_{sum} \geq \sum_{i=1}^{\ell} \log (k_i+1) - \epsilon$ since for  $ 1\leq i\leq \ell$, $R_i \geq \log (k_i+1) - \epsilon/ \ell$ and $R_{sum}=\sum_{i=1}^{\ell} R_i$. Hence, in order to maximize the sum-rate, our goal is to maximize the value of $\sum_{i=1}^{\ell} \log (k_i+1)$ given that $\sum_{i=1}^{\ell} k_i =t$. Assume that $t=k \ell+r$,
$r\in[\ell]$, then this maximum value will be achieved when choosing $k_1=\cdots =k_r=k+1$ and $k_{r+1}=\cdots =k_{\ell}=k$. The next corollary summarizes this result.  

\begin{corollary}\label{corEIPDU1}
For all $t$ and $\ell$, where $t=k \ell +r$, $r\in[\ell]$, 
\begin{align*}
\cR_{t,\ell}^{EIP:DU,z} & 
\hspace*{-0.05cm}
\geq r  \log (k+2)+(\ell-r)  \log (k+1) \\
& = \hspace*{-0.05cm}\ell\log\left(\left\lfloor\frac{t}{\ell}\right\rfloor\hspace*{-0.05cm}+\hspace*{-0.05cm}1\right) 
\hspace*{-0.05cm}+\hspace*{-0.05cm}
(t\bmod \ell)
\log\left(1\hspace*{-0.05cm}+\hspace*{-0.05cm}\frac{1}{\left\lfloor\frac{t}{\ell}\right\rfloor+1}\right).
\end{align*}
\end{corollary}
\begin{IEEEproof}
We choose $(k_1,\ldots,k_{\ell})$ such that $k_1=\cdots =k_r=k+1$ and $k_{r+1}=\cdots =k_{\ell}=k$ and thus $\sum_{i=1}^{\ell} k_i =t$. We note that $k= \left\lfloor\frac{t}{\ell}\right\rfloor$ and $r= t \mod \ell$. Since we presented in Construction \ref{constructEIPDUtl} an $[n,t,\ell;M_1,\ldots,M_t]^{EIP:DU,z}$ $\ell$-change $t$-write ELM code with sum-rate $R_{sum}=\sum_{i=1}^{\ell}R_i \geq  r  \log (k+2)+(\ell-r)  \log (k+1) -\epsilon$ for any $\epsilon >0$, we obtain the result in Corrollary \ref{corEIPDU1}. 
\end{IEEEproof}

From the above corollary, we have a lower bound of the maximum sum-rate of the EIP:DU model. Recall that $\cR^{EIP:DU,z}_{t,\ell} \leq \cR^{EIA}_{t,\ell} =\log \sum_{i=0}^{\ell} {t \choose i}$, that is the exact maximum sum-rate of the EIA model is an upper bound of the maximum sum-rate of the EIP:DU model. Hence, we obtain a lower bound and an upper bound of the maximum sum-rate of the EIP:DU ELM model. We note that when $t \leq \ell$, we always achieve the full capacity, that is, the maximum sum-rate is $t$. When $t > \ell$, the maximum sum-rate is difficult to compute exactly and there is a gap between the above lower and upper bounds. We illustrate the results for $\ell=2, t\in [3,25]$ in  the following figure.
\begin{center}
\begin{figure}[h]\label{fig1}
\centering
\includegraphics[width=6cm]{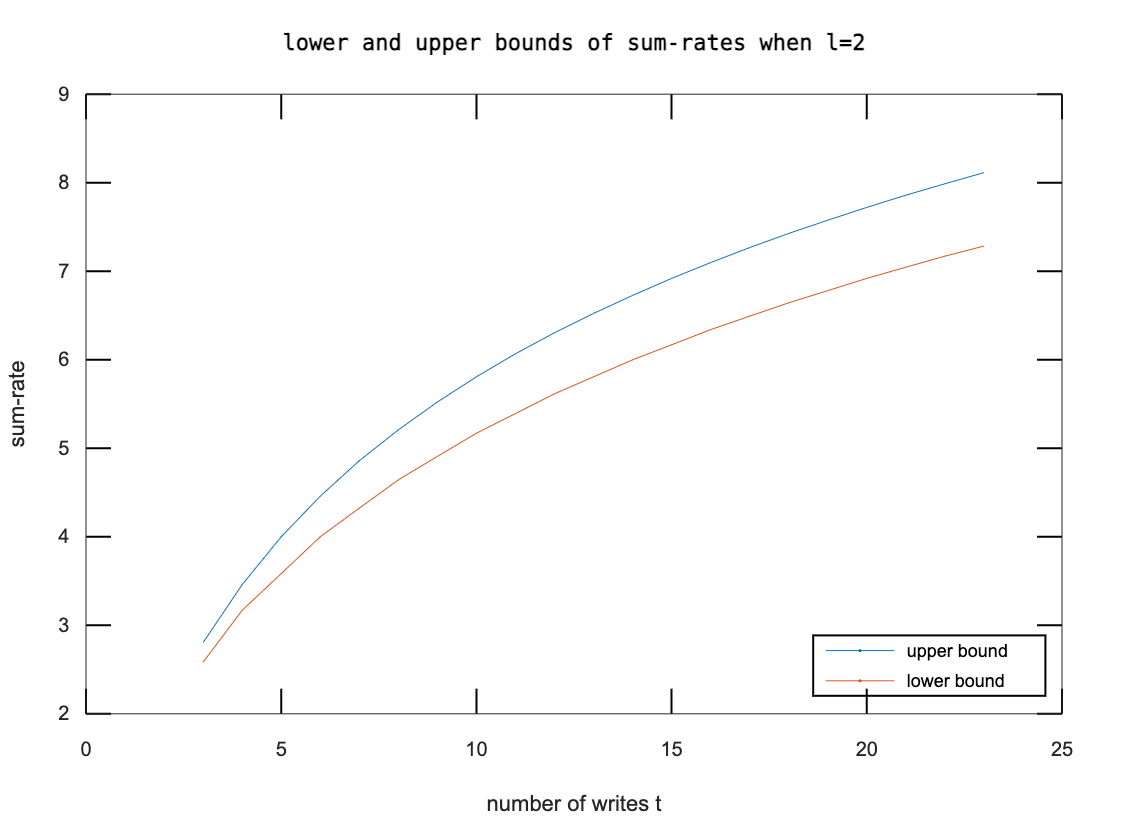}
\caption{The upper and lower bounds of the maximum sum-rates of the EIP:DU ELM codes when $\ell=2, t \in [3,25]$.}
\end{figure}
\end{center}
The following result shows that for $\ell=2$ the sum-rate of the ELM code from Construction \ref{constructEIPDUtl} is already close to the upper bound when $t$ is large and $n\to \infty$.
\begin{proposition}
For $\ell=2$ and $t\geq 3$, 
$\cR^{EIP:DU,z}_{t,2}\hspace*{-0.05cm} \geq \hspace*{-0.05cm}\cR^{EIA}_{t,2}\hspace*{-0.05cm}-\hspace*{-0.05cm}1$.
\end{proposition}

\begin{IEEEproof}
Recall that $\cR^{EIA}_{t,2} = \log \sum_{i=0}^{2} {t \choose i} = \log \frac{t^2+t+2}{2}.$
When $t$ is even, there exists a positive integer $t_1$ such that $t=2t_1$. In this case, 
\[\cR^{EIP:DU,z}_{t,2} \geq 2 \log (t_1 +1)=\log (t_1^2+2t_1+1)
\]
and,
\[\cR^{EIA}_{t,2}= 
\log \frac{4t_1^2+2t_1+1}{2}.
\]
Hence,
\[ \cR^{EIA}_{t,2} - \cR^{EIP:DU,z}_{t,2} \leq \log \frac{4t_1^2+2t_1+1}{2(t_1^2+2t_1+1)}\leq \log 2 =1.
\]
When $t$ is odd, there exists a positive integer $t_2$ such that $t=2t_2+1.$ In this case,
\[\cR^{EIP:DU,z}_{t,2} \geq \log (t_2 +1) 
+ \log (t_2+2)
=\log (t_2^2+3t_2+2)
\]
and,
\[\cR^{EIA}_{t,2}= 
\log \frac{4t_2^2+6t_2+4}{2}.
\]
Hence, \[ \cR^{EIA}_{t,2} - \cR^{EIP:DU,z}_{t,2} \leq \log \frac{4t_2^2+6t_2+4}{2(t_2^2+3t_2+2)} \leq \log 2 =1.
\]
In conclusion, the proposition is proven.
\end{IEEEproof}

We note that when $t=3$ and $\ell=2$, the maximum achievable sum-rate of the codes in Construction~\ref{constructEIPDUtl} is $\log 6 \approx 2.585$, while the upper bound is $\log 7 \approx 2.807.$ 
Lastly, we show how to improve this result for the $\epsilon$-error case.

The main ideas of the following construction are as follows. On the first two writes, we follow exactly the first two writes of Construction~\ref{constructEIA:DU32} which is a construction for a two-change three-write EIA:DU ELM code. After the second write, there are $\rho_1 n$ cells which were programmed twice, where $\rho_1 =p_{1,0} p_{2,1}.$ However, while the encoder in the EIA:DU model knows these positions, the encoder in the third write in the EIP:DU model does not know these positions. In order to overcome this difficulty, we use the following family of binary EU:DU WOM codes.
\begin{definition}\label{defqary}
An $[n,2;M_1,M_2]_2^{EU:DU,(p_{e_1},p_{e_2})}(p_1,p_2)$ two-write binary EU:DU WOM code is a coding scheme comprising of $n$ bits. It consists of two pairs of encoding and decoding maps $(\E_j^{EU:DU},\D_j^{EU:DU})$ for $ j = 1,2 $. For the map $\E_j^{EU:DU}$, $Im(\E_j^{EU:DU})$ is its image and $Im^*(\E_j^{EU:DU})$ is the set of all the cell-state vectors which can be obtained after the $j$-th write. We note that $Im(\E_0^{EU:DU})=Im^*(\E_0^{EU:DU})=\{(0,\ldots,0)\}$ and $Im^*(\E_2^{EU:DU})=\{\max\{\bfc_1,\bfc_2\} \text{ where } \bfc_i \in Im(\E_i^{EU:DU}): i=1,2\}$.
The encoding and decoding maps are defined as follows. 
For $j=1,2$, 
$$\E_j^{EU:DU}: [M_j] \to \B(n, (1-p_j,p_j))$$ and $$\D_j^{EU:DU}: Im^*(\E_j^{EU:DU}) \to [M_j]$$ such that
for all $m \in [M_j]$,
$$
\hspace{-18ex}\sum_{\hspace{17.5ex}(m,\bfc)\in [M_j]\times 
	Im^*(\cE_{j-1}^{EU:DU})}\hspace{-18.5ex}Pr(m)Pr(\bfc) 
	I_m\hspace{-0.5ex}\left(\cD_j^{EU:DU}(\hspace{-0.4ex} \max\{\bfc,\cE_j^{EU:DI}(m)\})\right)\hspace{-0.6ex} \leq \hspace{-0.5ex}
	p_{e_i}. 
$$
\end{definition}

Two-write binary EU:DU WOM codes have been studied for a long time \cite{WWZK84}. Recently,
in~\cite{HY17} several constructions of EU:DU WOM codes were presented. Assume that
there exists a capacity achieving code for the $Z$ channel, then the following result for EU:DU WOM codes can be received based upon the constructions from~\cite{HY17}. 

\begin{lemma}\label{lem.EUDU}\cite{HY17} 
For all $0 \le p_1,p_2\le 0.5$ and $\epsilon>0$ there exists an $[n,2;M_1, M_2]^{EU:DU,(0,\epsilon)}$ two-write binary EU:DU WOM code satisfying:
\begin{enumerate}[\hspace{-0.05cm}{\tiny \textbullet}\hspace{-0.07cm}]
\item 
$\bfc_1 \in \B(n,(1-p_1,p_1))$, and $R_1=\frac{\log M_1}{n} \geq h(p_1)-\epsilon.$
\item 
$\bfc_2 \in  \B(n,(1-p_2,p_2))$, and $R_2=\frac{\log M_2}{n} \geq h(p_1  p_2) - p_2  h(p_1)-\epsilon$,
\end{enumerate}
where $\bfc_i \in Im(\E_i^{EU:DU})$ for $i=1,2$.
\end{lemma}
We refer to the family of WOM codes from Lemma~\ref{lem.EUDU} as  
an $[n,2;M_1, M_2]_q^{EU:DU,(0,\epsilon)}(\epsilon, p_1,p_2)$ WOM code, where $M_1 = 2^{R_1n}$ and $M_2=2^{R_2n}$ are determined as the maximal possible values based on $\epsilon$, which tends to zero, and the probabilities $p_1$ and $p_2$.

We are now ready to present a construction of two-change three-write EIP:DU ELM code.
\begin{construction}\label{constructEIP:DU}
Given $p_{1,0},p_{2,0},p_{2,1}, p_3 \in [0,0.5]$,
we use the following two codes:
\begin{itemize}
\item An $[n,3,2;M_1,M_2,M'_3]^{EIA:DU,z}$ code from Construction \ref{constructEIA:DU32} with the first two pairs of encoder/decoder $(\E^{EIA:DU}_i,\D^{EIA:DU}_i)$ for $i=1,2.$
\item An $[n,2;M'_1,M_3]^{EU:DU,(0,\epsilon))}(\epsilon, \rho_1,p_3)$ two-write binary EU:DU WOM code from Lemma~\ref{lem.EUDU}, $\rho_1=p_{1,0} p_{2,1}$, with the pair of encoder/decoder in the second write $(\E_2^{EU:DU},\D_2^{EU:DU}).$
\end{itemize}
We construct an $[n,3,2;M_1,M_2,M_3]^{EIP:DU,(0,0,\epsilon)}$ two-change three-write EIP:DU ELM code where its 3 pairs of encoding/decoding maps $(\E^{EIP:DU}_j,\D^{EIP:DU}_j)$ for $j=1,2,3$ are defined as follows.
\begin{enumerate}[(1)]
\item 
For $i=1,2$, $\E^{EIP:DU}_{i}=\E^{EIA:DU}_i$ and $\D^{EIP:DU}_{i}=\D^{EIA:DU}_i$. That is, the first two writes of this EIP:DU ELM code  are exactly the same as the first two writes of the EIA:DU ELM code from Construction \ref{constructEIA:DU32}.
\item After the first two writes, we note that $ \rho_1  n$ cells are already programmed twice, and thus can not be programmed this time. Hence, we use the pair of encoder/decoder $(\E_2^{EU:DU},\D_2^{EU:DU})$ to encode/decode information. The pair of encoder/decoder in the third write is defined formally as follows:
\[\E^{EIP:DU}_{3}: [M_3] \times Im^*(\E^{EIP:DU}_2) \to [2]^n \]
such that for all $m_3 \in [M_3]$ and $\bfc_2 \in Im^*(\E^{EIP:DU}_2)$, $\E^{EIP:DU}_{3}(m_3,\bfc_2) =\overline{\E_2^{EU:DU}(m_3)}.$ 
Furthermore,
$$\D^{EIP:DU}_{3}: Im^*(\E_3^{EIP:DU}) \to [M_3]$$
such that for all $\bfc_3^* \in Im^*(\E_3^{EU:DU}),$
$\D^{EIP:DU}_{3}(\bfc_3^*)=\D^{EU:DU}_2(\overline{\bfc_3^*})=m_3.$
\end{enumerate}
On the first two writes, it is clear that $R_1\geq h(p_{1,0})-\epsilon$ and $R_2 \geq (1-p_{1,0}) h(p_{2,0})+p_{1,0} h(p_{2,1})-\epsilon$. In the third write, $R_3 \geq h(p_{1,0} p_{2,1}  p_3) - p_3  h(p_{1,0} p_{2,1})-\epsilon$.

In conclusion, we constructed a two-change three-write EIP:DU ELM code satisfying $R_1\geq h(p_{1,0})-\epsilon$, $R_2 \geq (1-p_{1,0}) h(p_{2,0})+p_{1,0} h(p_{2,1})-\epsilon$ and $R_3 \geq h(p_{1,0} p_{2,1} p_3)-p_3  h(p_{1,0}  p_{2,1})-\epsilon$ for all $\epsilon>0$.
\end{construction}

Therefore, the following region is achievable for the $\epsilon$-error case:
\begin{align*}
C^{EIP:DU}_{3,2}=\{&(R_1,R_2,R_3): R_1\leq h(p_{1,0}), \\
&R_2 \leq (1- p_{1,0}) h(p_{2,0})+p_{1,0}  h(p_{2,1}),\\
&R_3\leq h(p_{1,0} p_{2,1} p_3)-p_3  h(p_{1,0}  p_{2,1}),\\
&p_{1,0},p_{2,0},p_{2,1},p_3\in [0,1]\}.
\end{align*}
The sum-rate of the above code is $R_{sum}=R_1+R_2+R_3 \geq h(p_{1,0}) + (1-p_{1,0}) h(p_{2,0})+p_{1,0} h(p_{2,1}) + h(p_{1,0} p_{2,1} p_3)-p_3  h(p_{1,0}  p_{2,1}) - \epsilon$ for any $\epsilon >0$.
By choosing $p_{1,0}=3/7,p_{2,0}=1/2,p_{2,1}=2/3$, and $p_3=1/2$, we obtain the sum-rate $R_{sum}=R_1+R_2+R_3 \approx 2.64.$ 
\begin{remark}
In this section, we construct a family of zero-error $\ell$-change $t$-write EIP:DU ELM codes for any $\ell$ and $t$. Using some efficient encoding/decoding algorithms of the well-known binary $t$-write EI:DU WOM codes, we can encode/decode our EIP:DU ELM codes efficiently in polynomial time. When $n$ tends to infinity, we can obtain some codes with high sum-rate and thus get a lower bound on the maximal sum-rate of the EIP:DU model. We note that the lower bound is not tight even though it is close to the upper bound. We actually improve the lower bound for the $\epsilon$-error case when $\ell=2$ and $t=3$ in Construction \ref{constructEIP:DU}. Using some known polynomial time encoding/decoding algorithms of a two-write EU:DU WOM code in Lemma \ref{lem.EUDU}\cite{HY17}, the encoding and decoding algorithms in Construction \ref{constructEIP:DU} also run in polynomial time. Since the exact capacity region and the maximum sum-rate of the EIP:DU model are not known yet, we expect to have better constructions in near future.
\end{remark}

\section{Conclusion}\label{SEC.conclusion}
In this paper, we have proposed and studied a new coding scheme, called ELM codes. This family of codes can be used to increase the endurance of resistive memories by rewriting codes. This new family of rewriting codes generalizes the well-known WOM codes. We investigated the coding schemes of nine different models which depend upon the knowledge of the encoder and the decoder. In all these models, we focused on the capacity region and the achievable maximum sum-rate. In several important models, we also presented constructions of ELM codes with high sum-rate and some constructions of capacity-achieving codes. For future work, we are interested in practical constructions of capacity-achieving codes with efficient encoding/decoding algorithms, especially in the EIP:DU model.
\appendices
\section{}\label{app:proofs}
\begin{customthm}{\ref{thm:capacityEIP:DIA} - the converse part.}
The rates region $\widetilde{\cC}_{t,\ell}$ is a superset of the capacity region of $t$-write $\ell$-change ELM EIP:DIA model for the $\epsilon$-error case.
   That is,
    $
    \cC_{t,\ell}^{EIP:DIA,\epsilon}
  \subseteq 
  \widetilde{\cC}_{t,\ell}.
  $
\end{customthm}
\begin{IEEEproof}
Let $S_j$, $\hat{S}_j$, $V_{j}$, $1\le j \le t$, and $L$ be defined as in the proof of the converse part in Theorem~\ref{thm:capacityEIA}.
Thus, exactly as proved in Theorem~\ref{thm:capacityEIA}, we have $I(X_j;Y_j|V_{j-1}) \ge I(S_j;\hat{S}_j|V_{j-1})$, $I(S_j;\hat{S}_j|V_{j-1})\ge \log (M_j)-H(p_{e_j})-p_{e_j}\log (M_j)$, and
	\begin{equation*}
	\dfrac{1}{n}I(X_j;Y_j|V_{j-1}) \le
	\sum_{i=0}^{\ell-1}
	Pr(V_{j-1,L}=i)
	H(Y_{j,L}|V_{j-1,L}=i).
	\end{equation*}

Now, we set
	$p_{j,0}= Pr(X_{j,L}=1|V_{j-1,L} \mod 2 = 0)$
	and similarly
	$p_{j,1}= Pr(X_{j,L}=0|V_{j-1,L} \mod 2 = 1)$.
	Thus, for even $i<\ell$ $H(Y_{j,L}|V_{j-1,L}=i)=H(p_{j,0})$,
	and for odd $i<\ell$ $H(Y_{j,L}|V_{j-1,L}=i)=H(p_{j,1})$.
	We also define
	for $i\in [\ell+1]$
    $Q_{j,i}=Pr(V_{j,L}=i)$,
   and we note that $Q_{j,i}$ can be calculated as in
   Equation~(\ref{eq:EIP:DIAqProbFunc1}), and we use the notations $Q_{j,o}$ and $Q_{j,e}$ as defined above.
   Then,
    \begin{equation*}
	\begin{array}{ll}
	\hspace{-1ex}
	\dfrac{\log (M_j)}{n}	\hspace{-1ex} -	\hspace{-1ex}\epsilon_j	& 
	\hspace{-2ex}
	\le 	\hspace{-0.8ex}
	\dfrac{1}{n}I(X_j;Y_j|V_{j-1}) \\
	& \hspace{-2ex} \le  	\hspace{-0.8ex} \sum_{i=0}^{\ell-1} Pr(V_{j-1,L}=i)H(Y_{j,L}|V_{j-1,L}=i)\\
	&\hspace{-2ex} = 	\hspace{-0.8ex} \sum_{i=1}^{\ell/2} \left(Q_{j-1,2i-1}h\left(p_{j,1}\right)
	\hspace{-0.6ex}+
	\hspace{-0.6ex} Q_{j-1,2i-2}h\left(p_{j,0}\right)
	\right)\\
	&\hspace{-2ex} =	\hspace{-0.8ex}h\left(p_{j,1}\right)\sum_{i=1}^{\ell/2} Q_{j-1,2i-1}
	\hspace{-0.6ex}+
	\hspace{-0.6ex}h\left(p_{j,0}\right)\sum_{i=1}^{\ell/2} Q_{j-1,2i-2}\\
	& \hspace{-2ex} = 	\hspace{-0.8ex} Q_{j-1,o}
	h\left(p_{j,1}\right)
	+(Q_{j-1,e}-Q_{j-1,\ell})h\left(p_{j,0}\right),
	\end{array}
    \end{equation*}
	where $\epsilon_j=\frac{H(p_{e_j})+p_{e_j}\log(M_j)}{n}$,
	and the claim is implied.
\end{IEEEproof}

\begin{customthm}{\ref{thm:capacityEU:DIA}.}
The rates region $\overline{\cC}_{t,\ell}$ is the capacity region of $t$-write $\ell$-change ELM EU:DIA model for the $\epsilon$-error case. That is,
$\overline{\cC}_{t,\ell}= \cC_{t,\ell}^{EU:DIA,\epsilon}$.
\end{customthm}
\begin{IEEEproof}
To show the achievable region, we should prove
that for each $\epsilon>0$ and $(R_1, R_2, \ldots, R_t)\in \overline{\cC}_{t,\ell}$, there exists an \\ $[n,t;M_1,\ldots, M_t]_{t,\ell}^{EU:DIA, \bfp_e}$ ELM code,
where for all $1 \leq j \leq t$, $\frac{\log M_j}{n}\ge R_j-\epsilon$ and
$\bfp_e=(p_{e_1},\ldots,p_{e_t}) \leq (\epsilon,\ldots,\epsilon)$.
We use the well-known random channel-coding theorem \cite[p.~200]{CT91} on each write.
We describe the encoding and decoding on each write.

The $j$-th write presents a DMC with the input length-$n$ binary vector $X_j$ and the output is $(Z_{j-1},Y_j)$, where $Z_{j-1}\in [\ell+1]^n$ represents the times each cell was programmed before the $j$-th write, and $Y_j\in [2]^n$ represent the state of the memory after the $j$-th write.
Let $x_j=X_{j,k}$, $z_j=Z_{j-1,k}$, and $y_j=Y_{j,k}$ for some index $k$.
By the random coding theorem, for $n$ large enough, the following region is achievable
\begin{equation*}
\begin{array}{ll}
\hspace{-1ex} \Big\{ \hspace{-0.5ex} (R_1,\ldots,R_t)  | & \forall 1\leq j\leq t, R_j \leq I(x_j;y_j)\Big\}.
\end{array}
\end{equation*}

By the definitions and notations of the probabilities $p_{j'}$ and $Q_{j',i'}$,
\begin{align*}
    I(x_j;(z_{j-1},y_j))
    &{=}H(z_{j-1},y_{j})-H(z_{j-1},y_{j} \vert x_{j}) \\
    &=H(z_{j-1}) +H(y_{j}|z_{j-1})-H(z_{j-1},y_{j} \vert x_{j}) \\
    &\overset{(a)}{=}H(z_{j-1}) +H(y_{j}|z_{j-1})-H(z_{j-1}) \\
    &=H(y_{j}|z_{j-1})\\
    &=\sum_{i=0}^{\ell} Pr(z_{j-1}=i)H(y_{j}|z_{j-1}=i)\\
    &\overset{(b)}{=}\sum_{i=0}^{\ell-1} Pr(z_{j-1}=i)H(y_{j}|z_{j-1}=i)\\
    &=\sum_{i=1}^{\ell-1} Q_{j-1,i}h\left(p_{j}\right)\\
    &=\left(1-Q_{j-1,\ell}\right)h\left(p_{j}\right).
\end{align*}
Step $(a)$ follows from $H((z_{j-1},y_j)|x_i)=H(z_{j-1}|x_j)$ since $y_j$ is a function of $x_j,z_{j-1}$, and $H(z_{j-1}|x_j)=H(z_{j-1})$ because $z_{j-1}$ is independent on $x_j$. Step $(b)$ is implied by $H(y_j|z_{j-1}=\ell)=0$.
Hence, we can achieve the region $\widetilde{\cC}_{t,\ell}$ for the $\ell$-change $t$-write W$\ell$M EIP:DIA model for the $\epsilon$-error case.

The proof of the converse part is similar to the proof of this part in Theorem~\ref{thm:capacityEIA}. Let $S_j$, $\hat{S}_j$, $V_{j}$, $1\le j \le t$, and $L$ be defined as in the proof of the converse part in Theorem~\ref{thm:capacityEIA}.
Thus, exactly as proved in Theorem~\ref{thm:capacityEIA}, we have $I(X_j;Y_j|V_{j-1}) \ge I(S_j;\hat{S}_j|V_{j-1})$, $I(S_j;\hat{S}_j|V_{j-1})\ge \log (M_j)-H(p_{e_j})-p_{e_j}\log (M_j)$, and
	\begin{equation*}
	\dfrac{1}{n}I(X_j;Y_j|V_{j-1}) \le
	\sum_{i=0}^{\ell-1}
	Pr(V_{j-1,L}=i)
	H(Y_{j,L}|V_{j-1,L}=i).
	\end{equation*}

Now, we set
	$p_{j}= Pr(X_{j,L}=1)$.
	Thus, for $i<\ell$ $H(Y_{j,L}|V_{j-1,L}=i)=h(p_{j})$.
	We also define
	for $i\in [\ell+1]$
    $Q_{j,i}=Pr(V_{j,L}=i)$
   and we note that $Q_{j,i}$ can be calculated as in
   Equation~(\ref{eq:EIP:DIAqProbFunc1}). Then
    \begin{equation*}
	\begin{array}{ll}
	\hspace{-1ex}
	\dfrac{\log (M_j)}{n}-\epsilon_j	& 
	\hspace{-2ex}
	\le 
	\dfrac{1}{n}I(X_j;Y_j|V_{j-1}) \\
	& \hspace{-2ex} \le  \sum_{i=0}^{\ell-1} Pr(V_{j-1,L}=i)H(Y_{j,L}|V_{j-1,L}=i)\\
	&\hspace{-2ex} =\sum_{i=1}^{\ell-1} Q_{j-1,i}h\left(p_{j}\right) =\left( 1-Q_{j-1,\ell}\right) h\left(p_{j}\right), \\
	\end{array}
    \end{equation*}
	where $\epsilon_j=\frac{H(p_{e_j})+p_{e_j}\log(M_j)}{n}$,
	and the theorem is implied.
\end{IEEEproof}

\begin{customthm}{\ref{thm:comparing2}.}
For $t>\ell\ge 2$, 
$\cR^{EU:DIA,\epsilon}_{t,\ell} <
\cR^{EIP:DIA,\epsilon}_{t,\ell}$,
and hence 
$\cC^{EU:DIA,\epsilon}_{t,\ell} \subsetneq
\cC^{EIP:DIA, \epsilon}_{t,\ell}$.
\end{customthm}
\begin{IEEEproof}
Let $\overline{\bfR}=(\overline{R}_1,\overline{R}_2,\ldots, \overline{R}_t )$ be a rate tuple which achieves the maximum sum-rate $\cR^{EU:DIA,\epsilon}_{t,\ell}$,
and we denote by ${\overline{p}}_{j}$ and $\overline{Q}_{j,i}$, $1\le j\le t$ and $i\in[\ell+1]$, the probabilities which attain $\overline{\bfR}$ in $\overline{\cC}_{t,\ell}$.

Now we present a rate tuple $\widetilde{\bfR}=(\widetilde{R}_1,\widetilde{R}_2,\ldots, \widetilde{R}_t )\in {\widetilde{\cC}}_{t,\ell}> \overline{\bfR}$. Then, we conclude that $\widetilde{\bfR}\in \cC^{EIP:DIA,\epsilon}_{t,\ell}\setminus \cC^{EU:DIA,\epsilon}_{t,\ell}$,
which implies that $\cR^{EU:DIA,\epsilon}_{t,\ell} < \cR^{EIP:DIA,\epsilon}_{t,\ell}$
and 
$\cC^{EU:DIA,\epsilon}_{t,\ell} \subsetneq \cC^{EIP:DIA,\epsilon}_{t,\ell}$.

We assume now that $\ell$ is even, while the proof for the odd case is similar.
Since $\overline{\bfR}$ achieves maximum sum-rate we have
$\overline{p}_t=0.5$.
For all $j$ and $i$, $1\le j \le t-2$ and $i\in [\ell]$, we define $\widetilde{p}_{j,0}=\widetilde{p}_{j,1}=\overline{p}_j$.
In addition, $\widetilde{p}_{t-1,0}=0.5$, $\widetilde{p}_{t-1,1}=\overline{p}_{t-1}$,
and 
$\widetilde{p}_{t,0}=\widetilde{p}_{t,1}=0.5$.

Thus, for all $j$ and $i$, $1\le j \le t-2$ and $i\in [\ell]$, $\widetilde{R}_j=\overline{R}_j$ and $\widetilde{Q}_{j,i}=\overline{Q}_{j,i}$.
For the $(t-1)$-th write we have,
$\widetilde{R}_{t-1}= \overline{Q}_{t-2,o}h(\overline{p}_{t-1})+(\overline{Q}_{t-2,e}-\overline{Q}_{t-2,\ell})$ while 
 $\overline{R}_{t-1}=(1-\overline{Q}_{t-2,\ell})h(\overline{p}_{t-1})$,
and for the last write
$\widetilde{R}_{t}= \overline{R}_{t}=1-\overline{Q}_{t-1,\ell}$,

Now we prove that $\overline{p}_{t-1}<0.5$ which immediately implies that $\widetilde{R}_{t-1}>\overline{R}_{t-1}$ and thus completes the proof.
Recall that $\overline{R}_t=1-\overline{Q}_{t-1,\ell}=1-\overline{Q}_{t-2,\ell}-\overline{Q}_{t-2,\ell-1}\overline{p}_{t-1}$.
Thus, given the probabilities for the first $t-2$ writes, in order to achieve the maximal rate tuple $\overline{\bfR}$ we have to maximize $\overline{R}_{t-1}+\overline{R}_{t}$.
That is, we choose $\overline{p}_{t-1}$ which maximizes 
$(1-\overline{Q}_{t-2,\ell})h(\overline{p}_{t-1})-
\overline{Q}_{t-2,\ell-1}\overline{p}_{t-1}$.
The derivative is 
$(1-\overline{Q}_{t-2,\ell})\log ( \frac{1-\overline{p}_{t-1,1}}{\overline{p}_{t-1,1}})
-\overline{Q}_{t-2,\ell-1}$,
and the maximum is obtained for 
$\overline{p}_{t-1}=1 / (1+2^{{\overline{Q}_{t-2,\ell-1}}/({1-\overline{Q}_{t-2,\ell}})})$.
Since $\overline{\bfR}$ is maximal and $t>\ell\ge 2$, we have ${\overline{Q}}_{t-2,\ell-1}>0$, and therefore 
$\overline{p}_{t-1}\ne 0.5$.

\end{IEEEproof}

\end{document}